\documentclass[11pt]{article}
\usepackage[margin=1.1in]{geometry}
\usepackage{amsmath,amsthm}

\usepackage{amssymb}
\usepackage{graphicx}
\usepackage[colorlinks=true, allcolors=blue]{hyperref}
\usepackage{natbib}
\usepackage{subfig}
\usepackage{multirow}
\newcommand{\nothere}[1]{}
\newtheorem{assumption}{Assumption}
\newtheorem{theorem}{Theorem}

\title{\bf Efficient nonparametric estimators of discrimination measures with censored survival data}

\author{Marie Skov Breum \thanks{corresponding author: masb@sund.ku.dk} \ \textsuperscript{1} and Torben Martinussen\textsuperscript{1} \\
\footnotesize\textsuperscript{\textbf{1}} Section of Biostatistics, Department of Public Health, University of Copenhagen, Copenhagen, Denmark}

\begin{document}
	\maketitle
 	\begin{abstract}
		Discrimination measures such as the concordance index and the cumulative-dynamic time-dependent area under the ROC-curve (AUC) are widely used in the medical literature for evaluating the predictive accuracy of a scoring rule which relates a set of prognostic markers to the risk of experiencing a particular event.  Often the scoring rule being evaluated in terms of discriminatory ability is the linear predictor of a survival regression model such as the Cox proportional hazards model. This has the undesirable feature that the scoring rule depends on the censoring distribution when the model is misspecified. In this work we focus on linear scoring rules where the coefficient vector is a nonparametric estimand defined in the setting where there is no censoring. We propose so-called debiased  estimators of the aforementioned discrimination measures for this class of scoring rules. The proposed estimators make efficient use of the data and minimize bias by allowing for the use of data-adaptive methods for model fitting. Moreover, the estimators do not rely on correct specification of the censoring model to produce consistent estimation. We compare the estimators to existing methods in a simulation study, and we illustrate the method by an application to a brain cancer study.
	\end{abstract}

\noindent%

\section{Introduction}\label{sec:intro}

Evaluating the predictive accuracy of a set of prognostic markers is an important objective of many medical studies. For example, in the brain cancer study in Section \ref{sec:emprical} the objective is to evaluate whether a cheaper tumor growth biomarker predicts survival time not worse than a more expensive tumor growth marker when accounting for established genetic risk factors.
A popular metric for evaluating variable importance in a time-to-event setting is the so-called  concordance-index or c-index \citep{harrell1982evaluating, harrell1996multivariable, Heagerty2005, pencina2004overall}, which quantifies how well an estimated scoring rule discriminates between subjects with different event times. Considering two random subjects $(i,j)$ from the population of interest the  c-index is informally defined as
\begin{align}
    \label{eq:C}
    C=Pr(score(i) \geq score(j) \mid i \text{ has event before }j ),
\end{align}
where $\textit{score}(i)$ is the risk score for subject $i$.
An early estimator of the c-index for survival data was proposed by \cite{harrell1996multivariable}. However, as noted by \cite{GonenHeller2005} and \cite{Uno2011}, the limiting value of Harell's c-index estimator depends on the censoring distribution. To address this limitation, \cite{Uno2011} develop an estimator based on inverse probability of censoring weighting (IPCW), which is consistent for a time-truncated version of \eqref{eq:C} under independent censoring. 
\cite{Uno2011} use  the Cox model to construct their scoring and it  therefore depends on the censoring distribution if the model is not correctly specified.
\cite{Gerds2013concordance} propose an IPCW estimator that allows for marker-dependent censoring thus alleviating the independent censoring assumption of \cite{Uno2011}, but still requiring a correctly specified censoring model.
Although IPCW estimators are simple to compute, as they only require fitting a model for the probability of censoring, they are limited in  that they rely on correct specification of the censoring model to produce consistent estimation. Moreover, IPCW is generally not an efficient semiparametric method.
A different approach was taken by \cite{GonenHeller2005} who propose a plug-in estimator of the so-called concordance probability which is informally defined as
\begin{align}
   \label{eq:K}
   K=Pr(i \text{ has event before }j \mid score(i) \geq score(j)),
\end{align}
assuming that data are generated by a proportional hazards model. If this is not the case then their estimator is biased, and moreover it depends on the censoring distribution in the specific study which clearly is of no scientific interest. 
Note that when the event time and scoring rule are both continuous and we do not truncate the event-time, then the concordance measures in \eqref{eq:C} and \eqref{eq:K}  are identical. If, as is commonly done, we restrict to a specific time horizon $\tau$ (i.e, we only observe up to a given point in time) then the two measures are proportional.

As pointed out by \citep{blanche2019c} the c-index is not proper for the evaluation of $t$-year predicted risks, i.e. it may take a higher value for a scoring rule based on a misspecified model than for a scoring rule based on a correctly specified model. They recommended instead to use the so-called cumulative-dynamic time-dependent area under the ROC-curve ($\mbox{AUC}_t$) 
\citep{heagerty2000time,Heagerty2005,blanche2013review} that is proper in this scenario. The $\mbox{AUC}_t$ quantifies how well a scoring rule discriminates between subjects with different survival status at a given time point $t\leq \tau$, and is informally defined as
\begin{align}
    \label{eq:AUC_t}
    \mbox{AUC}_t = Pr(score(i) \geq score(j) \mid  i \text{ has event before }t \text{ and }j \text{ has event after }t ).
\end{align}
Frequently used estimators of the $\mbox{AUC}_t$ include plug-in estimators assuming that the data is generated by a proportional hazards model \citep{chambless2006estimation, song2008semiparametric} and IPCW estimators \citep{uno2007evaluating, hung2010optimal}. 

In this paper, we improve upon the foregoing approaches in several important directions.
By viewing  \eqref{eq:C}, \eqref{eq:K} and \eqref{eq:AUC_t}  as model-free  discrimination measures
we develop debiased estimators based on the corresponding efficient influence functions (eif), \citep{bickel1993efficient,kennedy2023semiparametric}. This give rise to efficient and flexible root-$n$ consistent estimators that are asymptotically model-free by enabling data-adaptive methods (e.g., machine learning). 
A further benefit is that the censoring mechanism can depend in an arbitrarily complex way on the considered markers and no specific working censoring model is required.
To the best of our knowledge, this is the first proposal of such a method for discrimination measures in a survival setting. None of the above mentioned methods enjoy these attractive properties. On the contrary, they all rely 
on some specific modeling  assumptions concerning either the conditional survival distribution and/or the censoring distribution. This may have undesirable consequences as it leads to bias and thus potentially flawed inference in  case the needed modeling assumptions are incorrect.

The three discrimination measures mentioned above all rely on a given scoring rule, which we denote by $\beta^T X$ with $X$ being the available markers that we wish to use to form the scoring rule. 
A standard choice is to use the Cox regression model to form the scoring rule using the Cox partial likelihood  estimator \citep{GonenHeller2005,Uno2011}.
If the Cox model is not correctly specified, however, then the Cox estimator converges to an estimand that depends on the censoring distribution, see \cite{Whitney2019} for a nice illustration of this, resulting in an ill defined scoring rule.
We take a different approach using  a well defined estimand, $\beta(P)$, that does not rely on 
data being generated from any specific model such as the Cox-model. 
To estimate $\beta(P)$ we first calculate its  efficient influence function and use the corresponding one-step estimator. This turns out to be crucial in order to obtain simple estimators for the three discrimination measures of interest.

This paper is organized as follows. In Section \ref{sec:prelim}, we introduce some notation and define the discrimination measures more formally. In Section \ref{sec:estimators}, we propose locally efficient non-parametric estimators of the truncated versions of discrimination measures based on the efficient influence function and we discuss estimation of $\beta(P)$. Section \ref{sec:largesamp} discusses the robustness properties of the proposed estimators. In Section \ref{sec:sim}, we conduct a simulation study to illustrate the large sample properties of the proposed estimators and competitors, and in Section \ref{sec:emprical} we apply the proposed novel estimators to some brain cancer data 
where it is of importance to judge whether a cheaper biomarker has a predictive power not worse than a more expensive marker. Some final remarks are provided in Section \ref{sec:conclucion}.  Technical derivations and further results are relegated to the Appendix.

\section{Preliminaries}\label{sec:prelim}

\subsection{Set-up and notation}

Let $T$ denote the continuous failure time  and $X$ the $d$-dimensional vector of markers. Let 
 $C$ denote the censoring time so that we observe $\tilde T = \min(T, C)$ and the censoring indicator $\Delta=I(T\leq C)$.  We assume that we observe data  in the time interval $[0,\tau]$ with $\tau<\infty$. Let $O=(\tilde T,\Delta,X)$ and suppose that we observe a sample of independent observations $O_1$,..., $O_n$ which are identically distributed according to some unknown probability distribution $P \in \mathcal{P}$ where $\mathcal{P}$ is a nonparametric statistical model. We use $\mathbb{P}_n\{f(O)\}=n^{-1}\sum_if(O_i)$ to denote the empirical measure  for some function $f$. Furthermore, we let $Pf = \int f(o) dP(o)$, and let $E$ denote expectations with respect to $P$. 
 
 Employing the standard notation for counting processes let  $N(t)=I(\tilde T\leq t,\Delta=1)$ denote the process that jumps when an event time of interest is observed and let $dM(t; X)$ be the corresponding martingale increment conditioning on $X$.  
We let $\lambda(t \mid x)$ denote the conditional hazard for the survival distribution given $X=x$ and $\Lambda(t\mid x)$ the corresponding cumulative hazard function. Furthermore, we let $S(t \mid x)$ denote the conditional survival function for $T$ and $K_C(t \mid x)$ the conditional survival function for $C$. We make the following assumption which ensures identifiability of the conditional distribution of $T$ given $X$.

 \begin{assumption}\label{assump:cens}
    $T$ and $C$ are conditionally independent given $X$.
 \end{assumption}

Let $\beta(P)$ be some estimand which reflects the association between survival and the markers, and define the  scoring rule
\begin{align}
\label{eq:score}
Y=\beta^T(P)X,
\end{align}
that we wish to evaluate in terms of predictive power. We return to how $\beta(P)$ can be chosen.

The following assumption will be needed when deriving the efficient influence function, which is the basis of constructing efficient estimators of the discrimination measures.
\begin{assumption}\label{assump:cov}
    We assume that the covariate vector $X$ contains at least one continuous covariate that we call $W$, and  write $X=(V,W)$ with $V$ denoting the remaining covariates. Also,
$$
\beta^TX=\beta_V^TV+\beta_W W,
$$
and assume $\beta_W \neq 0$. 
\end{assumption}

We let $\lambda_{T \mid Y}(t \mid y)$ denote the conditional hazard for the survival distribution given $Y=y$ and $\Lambda_{T \mid Y}(t \mid y)$ the cumulative hazard function. Let $S_{T \mid Y}(t \mid y)$ be the conditional survival function given $Y=y$.

\subsection{Discrimination measures}

The concordance index, also known as the c-index 
\citep{harrell1982evaluating,harrell1996multivariable,pencina2004overall, Uno2011,Heagerty2005}, for survival data is, in the truncated form, given by
\begin{align}
    \label{eq:C_tau}
    C_{\tau}(P) = P( Y_1 \geq Y_2\mid T_2 > T_1, T_1 \leq \tau ),
\end{align}
where $T_j$ is shorthand notation for a draw of the event time from the population where we restrict to $Y_j$.

A different concordance measure proposed by \cite{GonenHeller2005}  is 
\begin{align}
    \label{eq:K_tau}
    K_{\tau}(P) = P(T_2 > T_1, T_1 \leq \tau \mid Y_1 \geq Y_2).
\end{align}
 In \cite{GonenHeller2005}, the truncation by $\tau$ is left out and the resulting concordance probability is calculated assuming that data are generated by a Cox-model which further makes estimation possible. However, without the finite maximum follow-up time $\tau$, neither the concordance probability of Gönen and Heller nor the c-index, are in general identifiable. This is because it is not possible to assess the order of the event times that occur beyond the maximum follow-up time of the study. Moreover, the support of the censoring variable is usually shorter than the support of the survival time, which means that the tail part of the estimated survival function is unstable \citep{Uno2011}.

Clearly,  these two concordance measures are proportional to
\begin{align}
    \label{eq:Psi}
    \Psi_{\tau}(P) = P(T_2 > T_1, T_1 \leq \tau, Y_1 \geq Y_2),
\end{align}
in the considered setting, where both $Y$ and $T$ are continuous. Specifically,
$$
K_{\tau}(P)=2\Psi_{\tau}(P)\quad \mbox{and}\quad C_{\tau}(P) =2\Psi_{\tau}(P)/\left\{1-S^2(\tau) \right\},
$$
where $S(t)=P(T>t)$.
In the following we develop efficient nonparametric estimators of  both $K_{\tau}(P)$ and $C_{\tau}(P)$ mitigated by calculating the efficient influence function of $\Psi_{\tau}(P)$.

The c-index is widely used in practice, but it was recently argued  by \citep{blanche2019c} that the cumulative-dynamic time-dependent area under the ROC-curve ($\mbox{AUC}_t$) \citep{heagerty2000time,Heagerty2005,blanche2013review} 
defined for all $t\leq \tau$ by
\begin{align}
    \label{eq:AUC}
  \mbox{AUC}_t(P) = P( Y_1 \geq Y_2\mid T_1 \leq t, T_2 > t  ),
\end{align}
should be preferred if the aim is to predict risk of an event for a specific time horizon, $[0,t]$, say. 
For the same reason, we redefine $\beta(P)$ as $\beta_t(P)$.
Specifically, \citep{blanche2019c} showed that the c-index is not proper while the $\mbox{AUC}_t$ 
is proper. A proper discrimination measure takes its highest value when based on the true prediction model, see \citep{blanche2019c} for more details on proper discrimination measures.

We may define
\begin{align}
\label{eq:Theta}
\Theta_t(P)=P( Y_1 \geq Y_2, T_1 \leq t, T_2 > t  ),
\end{align}
so that $$ \mbox{AUC}_t(P)=\Theta_t(P)/[\{1-S(t)\}S(t)].$$ We will likewise derive the eif of $\Theta_t(P)$ and use it to develop efficient nonparametric estimators of $ \mbox{AUC}_t(P)$.

Going forward we write  $K_{\tau}$ for  $K_{\tau}(P)$ and likewise  for the other estimands unless we wish to stress the dependence on $P$.

\subsection{The estimand $\beta(P)$}
As mentioned earlier, an often used scoring rule is  $\check\beta^TX$, where $\check\beta$ is the Cox partial likelihood estimator. The partial likelihood estimator converges to a  well defined coefficient vector $\tilde\beta$ even if the Cox model is not correctly  specified, but $\tilde\beta$ has the undesirable feature that it depends on the censoring distribution in the actual study in the likely case where the Cox  model is misspecified. The scoring rule may thus reflect properties of the specific censoring distribution, which is of no scientific interest. We take a different approach defining the scoring rule in the setting where there is no censoring so it only reflects the association between survival and the markers. 

In this work, we will use the assumption-lean coefficient \citep{vansteelandt2022assumption, vansteelandt2024assumption} $\beta_t^{AL}(P)$ defined as
\begin{align}
   \label{eq:beta}
    \beta_t^{AL}(P)=\{\mbox{var}(X)\}^{-1}\mbox{cov}[g\{S(t\mid X)\},X],
\end{align}
where $g$ is some pre-specified link function. This estimand has the appealing property that if we take $g(x)=\log{\{-\log{(x)}\}}$ and, if in fact $S(t\mid X)=\exp{\{-\Lambda_0(t)e^{\theta^TX}\}}$ (ie the Cox model is correctly specified),  then $\beta_t^{AL}(P)=\theta$, but $\beta^{AL}_t(P)$ remains well defined otherwise.

Note however that the validity of our method does not rely on the specific choice of model-free estimand $\beta$ and others may be pursued as well.

\section{Estimation}\label{sec:estimators}
The efficient influence function  is a key object in  semiparametric efficiency theory as it characterizes the lower bound on the asymptotic variance of any regular estimators \citep{bickel1993efficient,vanderVaart2000, kennedy2023semiparametric}. Once the eif of an estimand has been derived, it can be used to construct estimators which enjoy desirable properties such as local efficiency, double robustness and root-$n$ consistency. There are several techniques for deriving efficient influence functions. In this paper we will use the Gâteaux derivative approach as detailed in \cite{ Hines2022demystifying} and \cite{kennedy2023semiparametric}. This approach is based on computing the pathwise derivative for a cleverly chosen submodel. An attractive property of Gateaux derivatives is that we can apply ordinary differentiation methods, such as the chain rule.

\subsection{Efficient nonparametric estimators of  $K_{\tau}$ and $C_{\tau}$}
The key to develop the novel estimators of the two concordance measures $K_{\tau}$ and $C_{\tau}$ is to calculate the efficient influence function of $ \Psi_{\tau}(P)$. 
Define
\begin{align}
\label{eq:h}
h_{\tau}(y_1,y_2)=\int_0^{\tau} S_{T \mid Y}(t\mid y_2)S_{T \mid Y}(t\mid y_1)d\Lambda_{T \mid Y}(t\mid y_1).
\end{align}
Then
\begin{align*}
\Psi_{\tau}(P)=\int\int_{y_1>y_2}  h_{\tau}(y_1,y_2)dF_Y(y_1)dF_Y(y_2),
\end{align*}
where $F_Y$ denotes the distribution function for $Y$. Note that under Assumption \ref{assump:cov} we have
\begin{align}
\label{eq:F_Y}
F_Y(y)=P(Y\leq y)=\int\int^{\frac{y-\beta_V^Tv}{\beta_W}}f_{W\mid V}(w \mid v)dwf_V(v)dv,
\end{align}
where $f_{W\mid V}$ is the conditional density of $W$ given $V$ and $f_V$ the density of $V$. 

In Appendix \ref{app:eif} we derive the eif of $ \Psi_{\tau}(P)$ by computing the Gateaux derivatives of \eqref{eq:h} and \eqref{eq:F_Y}, and applying the chain rule. This yields the following eif:

\begin{theorem} \label{theorem:eif1}
Under Assumptions \ref{assump:cens} and \ref{assump:cov} the efficient influence function of $\Psi_\tau$ is given by
    \begin{align}
\label{eif}
D^*_{\Psi_{\tau}}(O)(P)= g_{\tau}(X; P) + G_{\tau}(O, \dot\beta; P),
\end{align}
where 
\begin{align*}
g_{\tau}(X;P)=\int_{-\infty}^Y h_{\tau}(Y,y) dF_Y(y) + \int_Y^\infty h_{\tau}(y,Y) dF_Y(y)-2\Psi_\tau,
\end{align*}
and
\begin{align*}
 G_{\tau}(O,\dot\beta; P)=&-\int_Y^\infty\int_0^{\tau}S_{T \mid Y}(t\mid Y)\int_0^t\frac{dL_{\beta}(u;X)}{S_{T \mid Y}(u\mid Y)}S_{T \mid Y}(t\mid y)d\Lambda_{T \mid Y}(t\mid y)dF_Y(y) \\
 &-\int_{-\infty}^Y\int_0^{\tau} S_{T \mid Y}(t\mid y)S_{T \mid Y}(t\mid Y)\int_0^t\frac{dL_{\beta}(u;X)}{S_{T \mid Y}(u\mid Y)}d\Lambda_{T \mid Y}(t\mid Y) dF_Y(y) \\
&+\int_{-\infty}^Y\int_0^{\tau} S_{T \mid Y}(t\mid y)dL_{\beta}(t;X)dF_Y(y)+ \tilde H(\beta)\dot{\beta}(P),
\end{align*}
for
\begin{multline}
\label{eq:dL}
    dL_{\beta}(v;X)=\frac{dM(v;X)}{K_C(v\mid X)}-S(v\mid X)\int_0^v \frac{dM(u;X)}{S(u \mid X)K_C(u\mid X)}d\{\Lambda(v\mid X)-\Lambda_{T \mid Y}(v\mid Y)\}\\
    +S(v\mid X)d\{\Lambda(v\mid X)-\Lambda_{T \mid Y}(v\mid Y)\},  
\end{multline}
and $\tilde H(\beta)$ a non-stochastic constant given in \eqref{Constant_H}. Moreover, $\dot\beta(P)$ is the efficient influence function of $\beta$.    
\end{theorem}

In the next subsection, we describe how to use the eif in \eqref{eif} for estimation of $\Psi_{\tau}$. Before doing so we note that the constant $\tilde H(\beta)$  given in display
\eqref{Constant_H} in the Appendix has a complicated structure involving the density function of the covariate distribution. It is therefore desirable if the estimation procedure can avoid estimation of this constant.

\subsubsection{One-step estimator}
The one-step estimator \citep{vanderVaart2000,kennedy2023semiparametric} of $\Psi_{\tau}$ is defined as the solution to the estimating equation which sets set empirical mean of the eif $D^*_{\Psi_{\tau}}(O)(P)$ equal to zero.
This will obviously depend on unknown quantities, however. To this end we replace $\beta$ with some estimator $\hat\beta$, which we give in a moment, and $F_Y$ is replaced with its empirical counterpart. Since $K_{\tau}=2\Psi_{\tau}$, this results in the following  one-step estimator of the concordance probability $K_{\tau}$
\begin{equation}
\label{eq:os-K}
    \hat K_{\tau}=\frac{2}{n(n-1)} \sum_{i \neq j} I\left(\hat{\beta}^T X_i>\hat{\beta}^T X_j\right) \hat{h}_\tau\left(\hat{\beta}^T X_j, \hat{\beta}^T X_i\right)+\frac{1}{n}\sum_{i=1}^nG_{\tau}(O_i,\dot\beta(P_n); P_n).
\end{equation}
It is interesting to note that the first term in  \eqref{eq:os-K} corresponds to the estimator of \cite{GonenHeller2005} (GH-estimator) had we used the Cox partial likelihood estimator of $\beta$. The remaining part of  \eqref{eq:os-K}  is a debiasing term in case the data are not generated from a Cox model (given $X$). Note also
that we have used  the notation $P_n$ to indicate that we further need to replace all other unknown quantities, such as $S(\cdot \mid  X)$, by working estimates. 
We will be more specific about this later but leave it vague for now.
If data were in fact generated by a Cox model and we use the Cox partial likelihood estimator of $\beta$ then the debiasing term is negligible. In the more likely case where data are not generated by the Cox model the above one-step estimator can in principle be used to correct the bias of the GH-estimator. However, we do not recommend to use this procedure as it would require  estimation  of the constant $ \tilde H(\beta)$, which is not attractive due to the complicated structure
of this constant. However, this can be avoided all together if we further use the one-step estimator that solves the (empirical) version of the eif corresponding to the estimand $\beta(P)$.  Put in other words, by doing so, the term $ \tilde H(\beta)\dot{\beta}(P)$ can be dropped from the above eif. Since $C_{\tau}=K_{\tau}/\left\{1-S^2(\tau) \right\}$ this suggest the estimator
\begin{equation}
\label{eq:os-C}
\hat C_{\tau}=\hat K_{\tau}/\left [1-\{\hat S(\tau)\}^2 \right],
\end{equation}
where $\hat S(\tau)$ is its corresponding one-step estimator based on the observed data:
\begin{equation}
\label{Eff.KM}
\hat S(t)=n^{-1}\sum_iS_n(t\mid X_i)\left \{1-\int_0^{t}\frac{dM_n(u;X_i)}{S_n(u\mid X_i)K_C^n(u\mid X_i)}
\right\},    
\end{equation}
evaluated at $t=\tau$, and where $S_n(u\mid x)$ means an estimator of $S(u\mid x)$, and similarly with $dM_n(u;X_i)=dN_i(u)-I(u\leq \tilde T)d\Lambda_n(u\mid X_i)$ and  $K_C^n$.

\subsection{Estimation of $\beta(P)$}
It turns out to be crucial that the estimator of  $\beta$ satisfies the following assumption in order to obtain simple estimators for the three discrimination measures of interest.
\begin{assumption}\label{assump:AL}
    The estimator of $\beta$ solves the efficient influence curve equation, i.e.
    \begin{align*}
        \mathbb{P}_n \dot\beta(P_n) = o_P(n^{-1/2}),
    \end{align*}
    where $\dot\beta(P)$ is the efficient influence function of $\beta$. 
\end{assumption}

We now demonstrate how to construct an estimator of the assumption lean coefficient in \eqref{eq:beta} that satisfies Assumption \ref{assump:AL}. First note that the eif of \eqref{eq:beta} is given as follows.
\begin{theorem}
    Let $\beta_1=\mbox{cov}[g\{S(t\mid X)\},X]$ and $\beta_2=\mbox{var}(X)$. Under Assumption \ref{assump:cens} the efficient influence function of $\beta_t^{AL}$ is given by
    \begin{align*}
        D^*_{\beta}(O)(P)=\beta_2^{-1}D^{*}_{\beta_1}(O)(P)-\beta_2^{-1}D^{*}_{\beta_2}(O)(P)\beta^{AL}_t,
    \end{align*}
    where
    \begin{align*}
        D^{*}_{\beta_1}(O)(P)=&
\left[g\{S(t\mid X)\}-Eg\{S(t\mid X)\}\right](X-EX) -\beta_1 \\
&-(X-EX)g^{\prime}\{S(t\mid X)\}S(t\mid X)\int_0^t\frac{dM(u; X)}{K_C(u\mid X)S(u\mid X)},
    \end{align*}
    and
    \begin{align*}
        D^{*}_{\beta_2}(O)(P)= (X-EX)(X-EX)^T - \beta_2.
    \end{align*}
\end{theorem}

We can then construct the one-step estimator of $\beta_t^{AL}(P)$:
\begin{align}
 \label{eif.mult}
\hat \beta_t^{AL}=\{\hat{ \mbox{var}}(X)\}^{-1} \mathbb{P}_n\biggl ( &\left[g\{S_n(t\mid X)\}-\mathbb{P}_n g\{S_n(t\mid X)\}\right](X-\mathbb{P}_nX) \nonumber \\
&-(X-\mathbb{P}_nX)g^{\prime}\{S_n(t\mid X)\}S_n(t\mid X)\int_0^t\frac{dM_n(u; X)}{K_{C,n}(u\mid X)S_n(u\mid X)}\biggr ),
\end{align}
that solves the (empirical version) of the eif of $\beta_t^{AL}(P)$ equal to zero.

\subsection{Efficient nonparametric estimators of $\mbox{AUC}_t$}\label{sec:AUC}

Define 
$$v_t(y_1,y_2)=\{1-S_{T \mid Y}(t\mid y_1)\}S_{T \mid Y}(t\mid y_2),$$
and note that 
\begin{align*}
    \Theta_t(P)= \int \int_{y_1>y_2} v_t(y_1,y_2) dF_Y(y_1) dF_Y(y_2).
\end{align*}
The efficient influence function of $\Theta_t$ is given below.
\begin{theorem} \label{theorem:eif2}
 Under Assumptions \ref{assump:cens} and \ref{assump:cov} the efficient influence function of $\Theta_t$ is given by
$$
D^*_{\Theta_t}(O)(P)=\tilde g_t(X;P)+\tilde G_t(O,\dot\beta;P),
$$
where 
$$
\tilde g_t(X;P)=\int_{-\infty}^Y v_{t}(Y,y) dF_Y(y) + \int_Y^\infty v_{t}(y,Y) dF_Y(y)-2\Theta_t
$$
and
$$
\tilde G_t(O,\dot\beta;P)=\left\{ F_T(t) - F_Y(Y) \right\}S_{T \mid Y}(t \mid Y)\int_0^t\frac{dL_\beta(u; X)}{S_{T \mid Y}(u \mid Y)} +\check{H}(\beta)\dot{\beta}(P) 
$$
with $\check{H}(\beta)$ another constant given in \eqref{eq:Constant_H_AUC} and  $\dot\beta$ is the efficient influence function of $\beta$.     
\end{theorem}
This leads to  the following one-step estimator of $\Theta_t$ 
\begin{equation}
\label{os-theta}
    \hat \Theta_t=\frac{1}{n(n-1)} \sum_{i \neq j} I\bigl(\hat{\beta}^T X_i>\hat{\beta}^T X_j\bigr) \hat v_t\bigl(\hat{\beta}^T X_j, \hat{\beta}^T X_i\bigr)+\frac{1}{2n}\sum_{i=1}^n\tilde G_{t}(O_i,\dot\beta(P_n);P_n)
\end{equation}
where the part involving the complicated constant $\check{H}(\beta)$  can be dropped again as long as we use 
the estimator $\hat \beta_t$ which satisfies assumption \ref{assump:AL}. As we know how to estimate $S(t)$ using \eqref{Eff.KM} we thus propose the following estimator 
\begin{equation}
\label{eq:os-AUC}
\widehat{\mbox{AUC}}_t=\hat \Theta_t/[\{1-\hat S(t)\}\hat S(t)].
\end{equation}

\section{Large sample and robustness properties of proposed estimators}\label{sec:largesamp}
We will focus on the proposed estimator of  $\Psi_{\tau}$ but similar results apply for all estimators considered.
If we could correctly specify all the needed (working) models to calculate the proposed estimator $\hat K_{\tau}$ then it would have the eif given in \eqref{eif} as its influence function and in principle this could be used to estimate the asymptotic variance of the estimator by taking the empirical average of the squared influence function. 
We first turn, however, to the robustness properties of the estimator $\hat K_{\tau}$ focusing on robustness  in terms of consistency.

The dominating part of the remainder term for the proposed estimator of $\beta$ in terms of consistency is 
\begin{align*}
&-\left[ \mathbb{P}_ng\{S_n(t\mid X)\}-Eg\{S(t\mid X)\}\right ](EX-\mathbb{P}_nX)\\
&-E\left[ g^{\prime}\{S_n(t\mid X)\}S_n(t\mid X)(EX-\mathbb{P}_nX)\int_0^t\frac{\{K_{C,n}(u\mid X)-K_C(u\mid X)\}S(u\mid X)}{S_n(u\mid X)K_{C,n}(u\mid X)}d\{\Lambda_n-\Lambda\}(u\mid X)\right].
\end{align*}

The remainder term  corresponding to the proposed estimator of $\Psi_{\tau}$ with full data (ie without censoring) is
\begin{align*}
R_n^{\Psi_{\tau}}=-\int\int_{y_1>y_2} \{S_{T \mid Y,n}(t\mid y_2)-S_{T \mid Y}(t\mid y_2)\}   d\{F_{T \mid Y}^n(t\mid y_1)-F_{T \mid Y}(t\mid y_1)\}dF_Y(y_1)dF_Y(y_2)+o_p(n^{-1/2}),
\end{align*}
where we define $dF_{T \mid Y}(t\mid y)=S_{T \mid Y}(t\mid y)d\Lambda_{T \mid Y}(t\mid y)$. 
In the observed data case we get the same term plus some additional terms that all involve  
\begin{align*}
    dG_n(u\mid X)&=\biggl(\frac{K_{C,n}-K_C}{K_{C,n}}\biggr )(u\mid X)\{d\Lambda_n(u\mid X)-d\Lambda(u\mid X)\},
\end{align*}
see the Appendix for more details.
Hence, we can conclude, assuming the needed empirical quantities are estimated using estimator that converge at rate faster that $n^{1/4}$, that  the remainder term  is $o_p(n^{-1/2})$. Note also that if we were to use a working Cox-model for the distribution of $T$ given $X$, and it happens to be incorrectly specified, then the remainder term $R_n^{\Psi_{\tau}}$ is no longer negligible and bias would occur. This is confirmed in the numerical study reported in the next section. 

We thus get  that the  influence function is the efficient influence function, which in turn means that the proposed estimator is efficient.
However, we do not recommend to use the efficient influence function to estimate the variability of the proposed estimator as that would involve estimation of the covariates density function as the constant \eqref{Constant_H} would reappear. Instead we propose to use the nonparametric bootstrap approach to estimate the variability. This means that both estimation and inference can be carried out without having to estimate the density function of the covariates.

\section{Simulation studies}\label{sec:sim}

The aim of the simulation studies is to illustrate the theoretical properties of the proposed estimators.  In Section \ref{sec:sim1} we compare the proposed estimators to existing methods under different data-generating mechanisms and model misspecifications. In Section \ref{sec:sim2} we demonstrate the finite sample coverage of the nonparametric bootstrap confidence intervals. R-code to replicate the simulation studies is available in the supplementary materials.

\subsection{Simulation scenarios}
We simulate a vector of markers $X=(X_1, X_2)$ with $X_1 \sim \text{binom}(0.5)$ and $X_2\sim N(0, 1)$, and we consider the following data-generating schemes for the event times and the censoring times.
\begin{description}
    \item[Scenario 1:]
    \begin{itemize}
        \item[-] Event times are generated from a proportional hazards model with Weibull baseline hazard $\lambda(t \mid X)=k\lambda^{-1}(t/\lambda)^{k-1}\exp(\beta^TX)$ for $k=2$, $\lambda=0.1$ and $\beta = \{\text{log}(0.5), \text{log}(2)\}$.
        \item[-]  Censoring times are generated from a proportional hazards model with Weibull baseline hazard $\lambda_C(t \mid X_1)=k_C\lambda_C^{-1}(t/\lambda_C)^{k_C-1}\exp(\gamma^TX)$ with $k_C=1$, $\lambda_C=0.1$ and $\gamma=\{\text{log}(0.25), \text{log}(0.75)\}$. 
    \end{itemize}
        \item[Scenario 2:]
    \begin{itemize}
        \item[-] Event times are generated as in Scenario 1.
        \item[-]  Censoring times are generated from a proportional hazards model with Weibull baseline hazard $\lambda_C(t \mid X_1)=k_C\lambda_C^{-1}(t/\lambda_C)^{k_C-1}\exp(\gamma_1 X_1 + \gamma_2 X_2 + \gamma_3 X_1 X_2 + \gamma_4 X_2^2 + \gamma_5 X_1 X_2^2)$ with $k_C=1$, $\lambda_C=0.1$ and $\gamma=\{\log(1.5), \log(2), \log(0.25), \log(0.75), \log(0.25)\}$. 
    \end{itemize}
        \item[Scenario 3:]
    \begin{itemize}
        \item[-] Event times are generated from a  non-proportional hazards model with $\lambda(t \mid X)=(1-X_1) k_0\lambda^{-1}(t/\lambda)^{k_0-1}\exp(\beta_0 X_2) + X_1 k_1\lambda^{-1}(t/\lambda)^{k_1-1}\exp(\beta_1 X_2)  $ for $k_0=1$, $k_1=2$ and $\lambda=1$ and $\beta_0 = \text{log}(0.5)$, $\beta_1 = \text{log}(1)$.
        \item[-] Censoring times are generated from a proportional hazards model with exponential baseline hazard $\lambda_C \in \{0.2, 1.0, 1.5\}$. 
    \end{itemize}
\end{description}

We will compare the properties of the following estimators:
\begin{description}
    \item[GH:] A $\tau$-truncated version of the estimator proposed by \cite{GonenHeller2005} computed as
\begin{multline*}
    \frac{2}{n(n-1)}\sum_{i<j} \Bigg\{I(\check\beta^T X_i < \check\beta^T X_j) \frac{1-\exp\left\{-\hat{\Lambda}_0(\tau)(e^{\check\beta^T X_i}+e^{\check\beta^T X_j})\right\}}{1+\exp(\check\beta^T X_i-\check\beta^T X_j)} \\ + I(\check\beta^T X_i > \check\beta^T X_j) \frac{1-\exp\left\{-\hat{\Lambda}_0(\tau)(e^{\check\beta^T X_i}+e^{\check\beta^T X_j})\right\}}{1+\exp(\check\beta^T X_j-\check\beta^T X_i)} \Bigg\},
\end{multline*}
where $\check\beta$ is the Cox partial likelihood estimator and $\hat{\Lambda}_0$ the Breslow estimator of the baseline hazard.
    
    \item[IPCW-Uno:]  IPCW estimator by \cite{Uno2011, uno2007evaluating, hung2010optimal}  assuming marker-independent censoring. The estimator of the C-index is computed as
    \begin{align*}
        \frac{\sum_{i,j} I(\tilde T_i < \tilde T_j) N_i(\tau) I(\check\beta^T X_i > \check\beta^T X_j)\left\{\hat{G}(\tilde T_i) \right\}^{-2}}{\sum_{i,j} I(\tilde T_i < \tilde T_j) N_i(\tau)\left\{\hat{G}(\tilde T_i)\right\}^{-2}},
    \end{align*}
    and the estimator of the AUC as
    \begin{align*}
        \frac{\sum_{i,j} I(\tilde T_j> \tau)N_i(\tau) I(\check\beta^T X_i > \check\beta^T X_j)\left\{\hat{G}(\tilde T_i) \right\}^{-1}}{\left\{\sum_i I(\tilde T_i > \tau)\right\}\left\{\sum_i N_i(\tau) \hat{G}(\tilde T_i) ^{-1}\right\}},
    \end{align*}
    where $\check\beta$ is the Cox partial likelihood estimator and $\hat{G}(\cdot)$ is the Kaplan–Meier estimator for the censoring distribution $G(t)=P(C>t)$.

    \item[IPCW-Cox:] IPCW estimator allowing for marker-dependent censoring inspired by \cite{Gerds2013concordance, hung2010optimal,blanche2013review}. The estimator of the C-index is computed as
    \begin{align*}
        \frac{\sum_{i,j} I(\tilde T_i < \tilde T_j) N_i(\tau) I(\check\beta^T X_i > \check\beta^T X_j)\left\{\hat{K}_C(\tilde T_i \mid X_j )\hat{K}_C(\tilde T_i \mid X_i ) \right\}^{-1}}{\sum_{i,j}  I(\tilde T_i < \tilde T_j)N_i(\tau) \left\{\hat{K}_C(\tilde T_i \mid X_j )\hat{K}_C(\tilde T_i \mid X_i )\right\}^{-1}},
    \end{align*}
    and the estimator of the AUC as
    \begin{align*}
        \frac{\sum_{i,j} I(\tilde T_j> \tau)N_i(\tau) I(\check\beta^T X_i > \check\beta^T X_j)\left\{\hat{K}_C(\tilde T_i \mid X_i)\hat{K}_C(\tau \mid X_j) \right\}^{-1}}{\left\{\sum_i I(\tilde T_i > \tau) \hat{K}_C(\tau \mid X_i)\right\}\left\{\sum_i N_i(\tau) \hat{K}_C(\tilde T_i \mid X_i) ^{-1}\right\}},
    \end{align*}
     where $\check\beta$ is the Cox partial likelihood estimator and $\hat{K}_C(\cdot \mid X)$ is estimated using a Cox model.  
    
    \item[one-step-Cox:] One-step estimators proposed in Section \ref{sec:estimators} where the working models for the survival and censoring distributions are a Cox proportional hazards regression adjusted for main effects of $X_1$ and $X_2$. The Cox proportional hazards  regressions were implemented using the \texttt{cox.aalen} function from the \texttt{timereg} package \citep{timereg-package, ScheikeMartinussenbook}.
    \item[one-step-RF:] One-step estimators proposed in Section \ref{sec:estimators} where the working models for the survival and censoring distributions are survival random forest regressions implemented using the \texttt{survival\_forest} function from the \texttt{grf} package \citep{grf-package}. We used default values of the tuning parameters.
    \item[one-step-RF-CF:] as above with 5-fold cross-fitting 
\end{description}

 Note that for the GH, IPCW-Uno and IPCW-Cox estimators the scoring rule is the linear predictor of a Cox PH model. For the one-step estimators, the scoring rule is constructed using the assumption lean coefficient in \eqref{eq:beta}. In Scenarios 1 and 2 where data is generated from a proportional hazards model these two scoring rules coincide, but not in Scenario 3.  

\subsection{Simulation study 1: comparison with existing methods} \label{sec:sim1}

\subsubsection{Robustness properties}
\begin{figure}[h!]
    \centering
    \includegraphics[scale=.61]{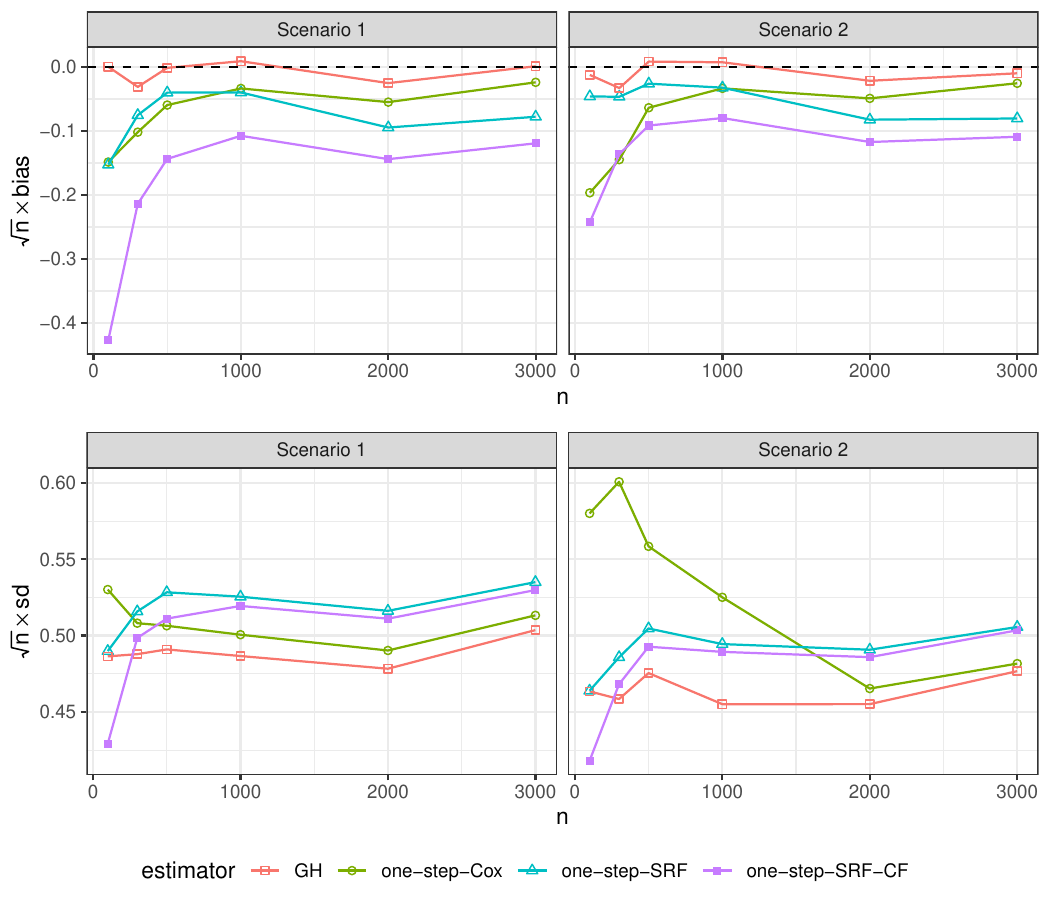}
    \caption{Simulation results for $K_\tau$. Row 1 is bias times $\sqrt{n}$ and row 2 is empirical standard deviation times $\sqrt{n}$.}
    \label{fig:simK}
\end{figure}

\begin{figure}[h!]
    \centering
    \includegraphics[scale=.61]{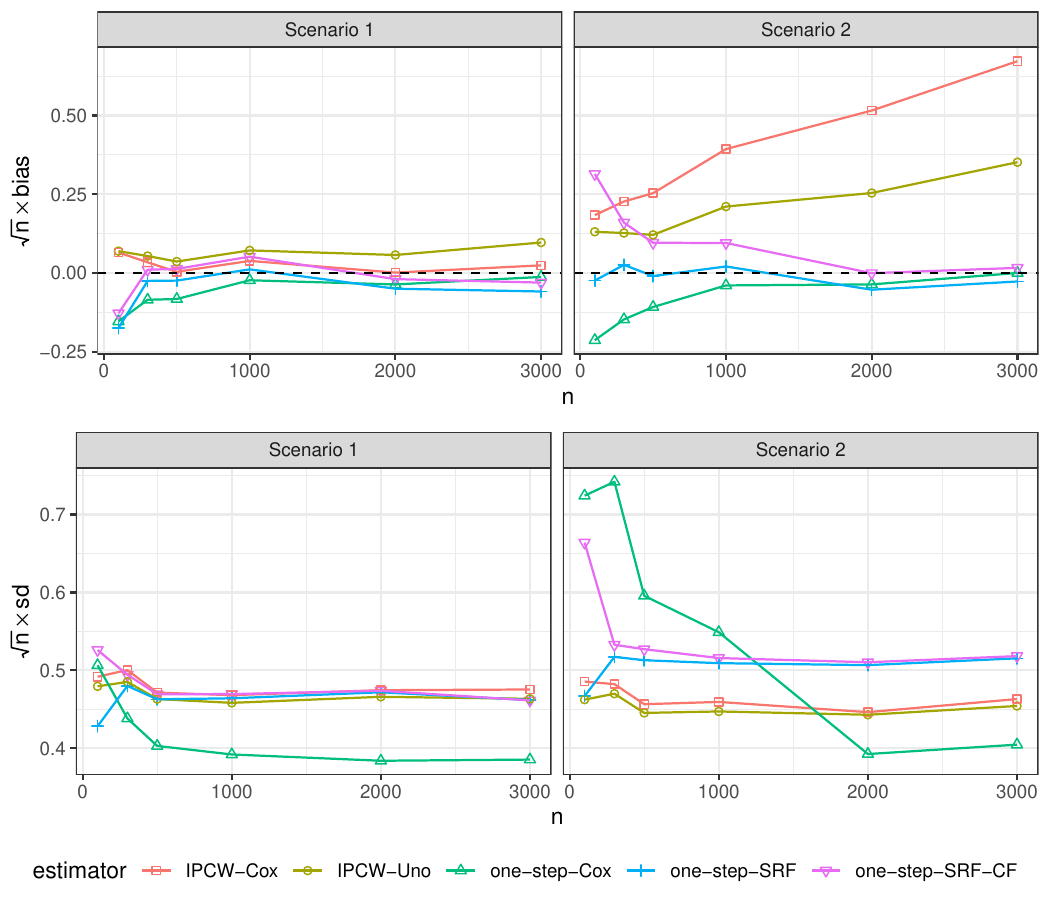}
    \caption{Simulation results for $C_\tau$. Row 1 is bias times $\sqrt{n}$ and row 2 is empirical standard deviation times $\sqrt{n}$.}
    \label{fig:simC}
\end{figure}

\begin{figure}[h!]
    \centering
    \includegraphics[scale=.61]{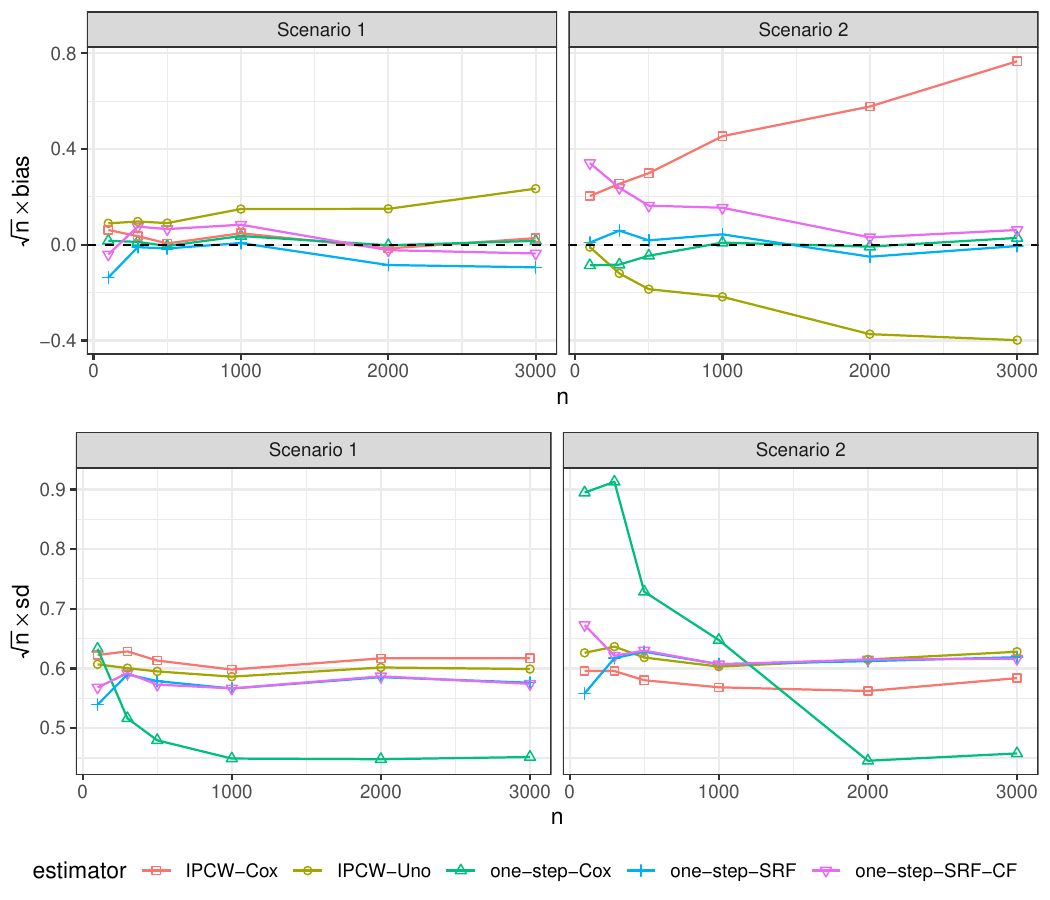}
    \caption{Simulation results for $\mbox{AUC}_\tau$. Row 1 is bias times $\sqrt{n}$ and row 2 is empirical standard deviation times $\sqrt{n}$.}
    \label{fig:simAUC}
\end{figure}

We first consider the data-generating mechanisms described in Scenarios 1 and 2. We set $\tau=t=8$, and we note that the true values of the discrimination measures are $K_{\tau, 0}=0.446$, $C_{\tau, 0}=0.697$ and $\mbox{AUC}_{t, 0}=0.745$. For each of the  data-generating mechanisms we simulate data with a sample size of $n \in \{100, 300, 500, 1000, 2000, 3000\}$.
We repeated this 1000 times and we computed the absolute bias scaled by $\sqrt{n}$ and the empirical standard deviation (sd) scaled by $\sqrt{n}$. The results are displayed in Figures  \ref{fig:simK} - \ref{fig:simAUC}.  

We see that, as expected, the GH and IPCW-Cox estimators perform well in Scenario 1 where data is generated from a Cox proportional hazards model and the censoring model is correctly specified. The IPCW-Uno estimator is biased across both scenarios as the marker-independent censoring assumption is violated, and when the censoring model is misspecified as in Scenario 2 the scaled bias of the IPCW-Cox does not tend to zero.  As expected given the large sample properties in Section \ref{sec:largesamp} the one-step-Cox estimator is robust to misspecification of the censoring model in Scenario 2. Importantly, the simulation studies demonstrate that the scaled bias of one-step estimator based on survival random forest stabilizes to a small value across both scenarios. The bias of the cross-fitted estimator one-step-SRF-CF is generally larger than the bias of the non cross-fitted estimator one-step-SRF for the smaller sample sizes, but tends to zero faster than $1/\sqrt{n}$.

\newpage
\subsubsection{Non-proportional hazards}

Next we simulate data from the data-generating mechanism described in Scenario 3  with a sample size of $n \in \{100, 300, 500, 1000, 2000\}$. We set $\tau=t=1$, and we note that the true values of the discrimination measures are $K_{\tau, 0}=0.515$, $C_{\tau, 0}=0.598$ and $\mbox{AUC}_{t, 0}=0.632$. 

\begin{figure}[h!]
    \centering
    \includegraphics[scale=.7]{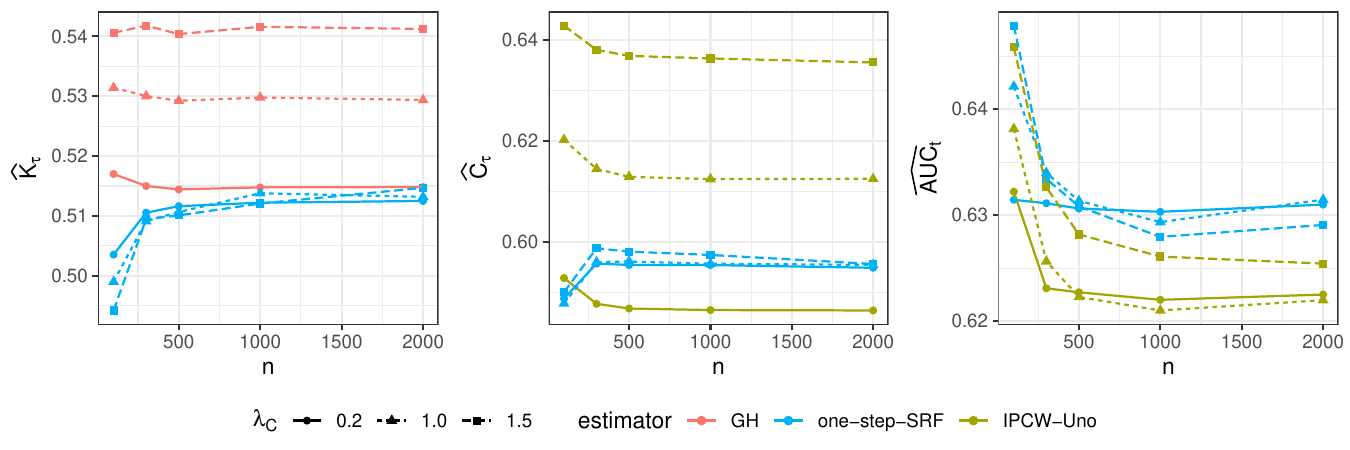}
    \caption{Simulation results under Scenario 3 with $\lambda_C \in \{0.2, 1.0, 1.5\}$. Left plot is the average value of estimators of $K_\tau$; middle plot is the average value of estimators of $C_\tau$; right plot is the average value of estimators of $\mbox{AUC}_t$.}
    \label{fig:sim3}
\end{figure}

For each simulated data set we compute the GH estimator and the one-step-SRF estimator of $K_\tau$, and the IPCW-Uno and one-step-SRF estimators of $C_\tau$ and $\mbox{AUC}_t$. In Figure \ref{fig:sim3}, we display the average value of the estimators based on $1000$ simulated data sets. 

\noindent
Unsurprisingly, we see that the value of the GH and IPCW-Uno estimators depend on the censoring distribution, while the value of the one-step-SRF estimator does not. The results for the one-step-Cox is given in the Supplementary Material. This estimator is not protected against misspecification of the survival model and some bias is also seen in that figure.

\subsection{Simulation study 2: finite sample coverage} \label{sec:sim2}
We simulate data from the data generating mechanism in Scenarios 1,2 and 3 with a sample size of $n \in \{100, 300, 500\}$. In Scenario 3 we set $\lambda_C=1.0$, and as above we set $\tau=t=8$ in Scenarios 1 and 2, and $\tau=t=1$ in Scenario 3.

For each simulated data set we compute the one-step-Cox estimators (Table \ref{tab:simPropHaz}) and the one-step-SRF estimators (Table \ref{tab:simRF}) of the concordance probability $\hat{K}_{\tau}$, the c-index $\hat{C}_{\tau}$ and the $\mbox{AUC}_t$. This was repeated 1000 times and we computed the bias, empirical standard deviation (sd), the bootstrap standard error based on 500 replications (se) and the coverage of a 95\% Wald-type confidence interval (cov). 

\begin{table}[h!]
\centering
\caption{Simulation results for one-step-Cox. bias is the average bias across simulations ($\times 100$); sd is the empirical standard deviation ($\times 100$); se is the average standard error across simulations ($\times 100$); cov is the coverage of a 95\% Wald type confidence interval  ($\times 100$). Each entry is based on 1000 replicates.}
\begin{tabular}{llrrrrrrrrrrr}
\hline
& & \multicolumn{3}{c}{Scenario 1} & & \multicolumn{3}{c}{Scenario 2} & & \multicolumn{3}{c}{ Scenario 3} \\  \cline{3-5} \cline{7-9}  \cline{11-13} 
                              & \multicolumn{1}{r}{n}  &  100      & 300     & 500  &  & 100      & 300     & 500   &  & 100      & 300     & 500           \\ \hline
\multirow{3}{*}{$\hat{K}_{\tau}$}   & bias &   -1.48     & -0.59   &  -0.27   & & -1.96 & -0.84& -0.29& & -1.78& -0.58 & -0.27        \\
                             & sd   & 5.30      &  2.93  & 2.26     & & 5.80& 3.47 & 2.50 & & 5.09& 3.16&  2.34        \\
                               & se  & 5.37       &  3.04 &  2.32   & & 5.61& 3.44 & 2.54 & & 5.34& 3.30&  2.49       \\
                               & cov  &  94.6      &  96.1   &  95.4 &  & 94.3& 96.8& 94.8 & & 94.2 & 95.3& 95.3           \\[.3cm]     
\multirow{3}{*}{$\hat{C}_{\tau}$}   & bias &   -1.53     & -0.49    & -0.37      & &-2.13 & -0.85& -0.48& & -1.49& -0.46& -0.25      \\
                              & sd   &  5.07     &  2.52   & 1.80     &  & 7.24& 4.28 & 2.66 & &5.45& 3.33&  2.45     \\
                               & se  &  5.14      & 2.66   &  1.96   & & 6.36&  3.83 & 2.71 & &5.74& 3.49& 2.61  \\
                               & cov  & 94.8      &  95.5  &  95.7   & & 94.5& 95.5& 95.6& &92.6& 96.0&   95.5     \\[.3cm]   
\multirow{3}{*}{$\widehat{\mbox{AUC}}_t$} & bias & 0.17      & 0.07  & -0.02        &   & -0.86 & -0.48& -0.21 & & 2.22 & 1.38& 1.26  \\
                              & sd   & 6.33       & 2.98    & 2.14        &    & 8.95 & 5.27 & 3.26& &7.14& 4.29& 3.08     \\
                               & se  & 6.59       & 3.17   & 2.32       &    & 8.48 & 4.57& 3.22 & &12.4& 4.47& 3.32     \\
                               & cov  & 94.2       & 94.7   &  96.7     &    & 93.2 & 95.0& 96.5 & &92.5& 92.5& 92.7     \\
                              \hline    
\end{tabular}
\label{tab:simPropHaz}
\end{table}

In Table \ref{tab:simPropHaz} we see that, unsurprisingly, when the working models are estimated using a correctly specified Cox model in Scenario 1 the one-step estimators are unbiased and the average estimated standard error closely approximates the empirical standard deviation. The Wald-type 95\% confidence interval exhibits a coverage close to the nominal level. The estimators remain unbiased in Scenario 2 where the censoring model is misspecified, and moreover the average estimated standard error closely approximates the empirical standard deviation and the confidence interval exhibits a coverage close to the nominal level. In Scenario 3 where the Cox PH assumption does not hold, the estimators exhibit a large bias (compared to the one-step-SRF) and the coverage of the confidence intervals for the $\mbox{AUC}_t$ is anti-conservative.

\begin{table}[h!]
\centering
\caption{Simulation results for one-step-SRF. bias is the average bias across simulations ($\times 100$); sd is the empirical standard deviation ($\times 100$); se is the average standard error across simulations ($\times 100$); cov is the coverage of a 95\% Wald type confidence interval  ($\times 100$). Each entry is based on 1000 replicates.}
\begin{tabular}{llrrrrrrrrrrr}
\hline
& & \multicolumn{3}{c}{Scenario 1} & & \multicolumn{3}{c}{Scenario 2} & & \multicolumn{3}{c}{ Scenario 3} \\  \cline{3-5} \cline{7-9}  \cline{11-13} 
                              & \multicolumn{1}{r}{n}  &  100      & 300     & 500  & &  100      & 300     & 500   & &  100      & 300     & 500           \\ \hline
\multirow{3}{*}{$\hat{K}_{\tau}$}   & bias &   -1.52     & -0.43  &  -0.18    &   &  -0.45& -0.27& -0.12 & & -1.39 & -0.38& -0.25    \\
                             & sd   & 4.90      &  2.98  & 2.36        & & 4.64 & 2.80 & 2.26& &4.58& 3.20 & 2.35   \\
                               & se  & 4.89       &  2.98  &  2.39      &     & 4.71& 2.82 & 2.19& & 9.86 &  9.90& 8.86\\
                               & cov  &  94.1      &  94.8   &  95.3     &      & 95.1 & 95.6 & 93.8 & & 95.2& 96.7& 97.5 \\[.3cm]    
\multirow{3}{*}{$\hat{C}_{\tau}$}   & bias &   -1.74     & -0.14    & -0.11  &  & -0.21 & 0.14 & -0.04 & & -0.76& 0.09 &  0.06    \\
                              & sd   &  4.27     &  2.76  & 2.07       &      & 4.68& 2.98& 2.30 & & 4.83& 3.43& 2.49 \\
                               & se  &  4.35      & 2.71   &  2.23   &    & 4.75& 2.88& 2.28& & 11.6 &  11.1& 9.80\\
                               & cov  & 93.2     &  93.9  &  95.0    &     & 94.2& 93.7& 94.0& &94.5& 96.8 &  98.0 \\[.3cm]   
\multirow{3}{*}{$\widehat{\mbox{AUC}}_t$} & bias &  -1.37     & -0.05 & -0.07       &  & 0.13 & 0.33& 0.08& &
1.28& 0.48& 0.18    \\
                              & sd   &  5.38      & 3.40   & 2.58     &      & 5.59 & 3.56& 2.82& & 6.46& 4.40& 3.44  \\
                               & se  &  5.39     & 3.33   &  2.60       &   & 5.66 & 3.45& 2.73& & 17.8& 4.43&  3.45 \\
                               & cov  & 94.3      & 93.9   &  95.5     &    & 93.7 & 91.9& 93.5& & 94.8& 93.2& 93.9    \\
                              \hline  
\end{tabular}
\label{tab:simRF}
\end{table}

In Table \ref{tab:simRF} we see that when the working models are estimated using survival random forest the estimators are unbiased across all 3 scenarios In scenarios 1 and 2 the average estimated standard error closely approximates the empirical standard deviation and the confidence interval exhibits a coverage close to the nominal level, though slightly anti-conservative for the smaller sample size ($n=100$). In scenario 3 the coverage is conservative for $K_\tau$ and $C_\tau$ but close to the nominal level for $\mbox{AUC}_t$.

\section{Brain Tumor Data}\label{sec:emprical}

\begin{figure}[h!]
    \centering
    \subfloat[]{\includegraphics[scale=.4]{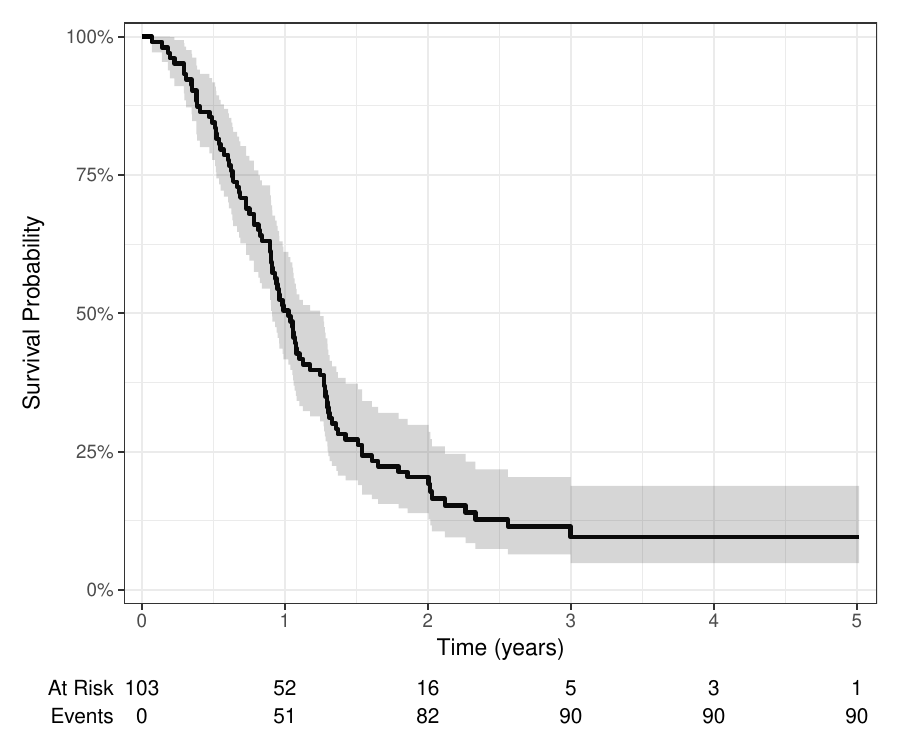}}
    \qquad
    \subfloat[]{\includegraphics[scale=.4]{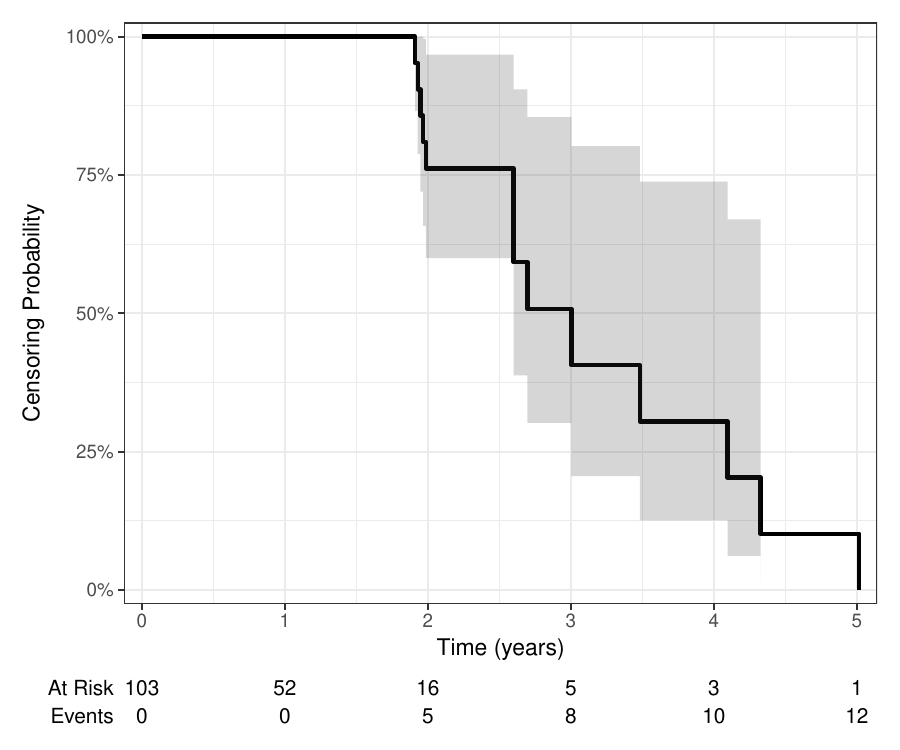}}
    \caption{Brain tumor data. Kaplan-Meier estimates of (a) the survival probability and (b) the censoring probability. The numbers at risk and  cumulative number of events are given below.}
    \label{fig:KM1}
\end{figure}

This study consists of data from a  total of 103 glioma patients. At first suspicion of tumor growth a  simultaneous PET/MRI  brain scan was made for each patient. 
Tumor volume determined by  the PET and MRI is called FET-VOL and BV-VOL, respectively. 
Besides this, there is also information about a binary genetic marker `methyl' which is a well established predictor. 
Figure \ref{fig:KM1} shows the Kaplan-Meier curves of the survival probability and the censoring probability, as well as the number of people at risk and the cumulative number of events. 
\noindent
The primary aim of the study was to investigate  whether the cheaper biomarker  BV-VOL is as good as the more expensive
biomarker FET-VOL in terms of discriminatory ability concerning survival within the first to  two year after suspicion of tumor  growth.  We therefore  compare the discriminatory power of the  scoring rules based on (A) FET-VOL and methyl; and (B) BV-VOL and methyl.

Table \ref{tab:oneStep-brain} shows the estimated estimated $C_{\tau}$ and $K_{\tau}$ for $\tau=2$ years using scoring rules (A) and (B), and for each of these using Cox proportional hazards regression and survival random forest regression, respectively, to estimate the needed predicted survival and censoring probabilities. Table \ref{tab:estimates-brain} shows the Gönen and Heller estimator (GH), the IPCW estimator of the c-index of \cite{Uno2011} and the IPCW estimator of the c-index of \cite{Gerds2013concordance} where the censoring model is a Cox PH model adjusted for the markers in (A) and (B). The scoring rule used for the estimators in Table \ref{tab:estimates-brain} is the linear predictor of a Cox PH regression model. 
\begin{table}[h!]
\centering
\caption{Brain tumor data. One step estimates ($\times 100$) of $\mbox{C}_{\tau}$ and $\mbox{K}_{\tau}$ for $\tau=2$ with 95\% confidence intervals.  }
\label{tab:oneStep-brain}
\begin{tabular}{llllll}
\hline
                                   & \multicolumn{2}{c}{Cox PH}                                &  & \multicolumn{2}{c}{SRF}                                  \\ \cline{2-3} \cline{5-6} 
                                   & \multicolumn{1}{c}{score A} & \multicolumn{1}{c}{score B} &  & \multicolumn{1}{c}{score A} & \multicolumn{1}{c}{score B} \\ \hline
\multirow{2}{*}{$\hat{\mbox{C}}_{\tau}$} & 64.1                        & 68.6                        &  & 65.0                        & 68.6                            \\
                                   & (54.5; 73.6)                & (64.0; 73.3)                &  & (58.9; 71.1)                & (64.0; 73.3)            \\
\multirow{2}{*}{$\hat{\mbox{K}}_{\tau}$} & 61.3                        & 65.7                     &  & 62.1                        & 65.6                      \\
                                   & (52.3; 70.3)                & (61.4; 70.0)                &  & (56.2; 68.0)                & (61.5; 69.8)              \\ \hline    
\end{tabular}
\end{table}

\begin{table}[h!]
\centering
\caption{Brain tumor data. GH and IPCW estimator ($\times 100$) of $\mbox{C}_{\tau}$ and $\mbox{K}_{\tau}$ for $\tau=2$ with 95\% confidence intervals. The estimators are based on Cox PH working models and a Cox PH scoring rule.}
\label{tab:estimates-brain}
\begin{tabular}{lll}
\hline
                                   & \multicolumn{1}{c}{score A} & \multicolumn{1}{c}{score B}  \\ \hline
GH &     60.6 (56.3;64.9)                   &    66.1 (62.1; 70.1)               \\
IPCW-Uno &  63.0 (57.4;68.6)                 &    65.9 (60.6; 71.3)               \\
IPCW-Cox &  67.8 (62.7; 73.0)                &   71.6 (66.8; 76.5)                 \\ \hline    
\end{tabular}
\end{table}

\begin{figure}[h!]
    \centering
    \includegraphics[scale=.8]{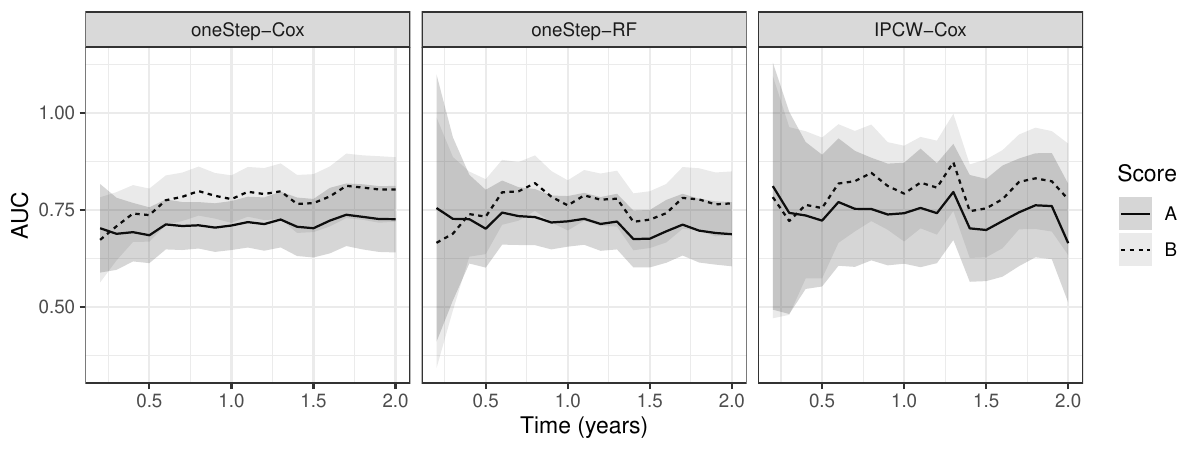}
    \caption{Brain tumor data. One-step estimates of $\mbox{AUC}_t$ versus time with pointwise 95\% confidence intervals.}
    \label{fig:AUC}
\end{figure}

Figure \ref{fig:AUC} shows the estimated $\mbox{AUC}_t$ as a function of time for $t\leq \tau$ using scoring rules (A) and (B). As before estimation of the  predicted survival and censoring probabilities is based on Cox proportional hazards regression and survival random forest regression, respectively. Also shown in Table \ref{tab:estimates-brain} and  Figure \ref{fig:AUC} are the pointwise 95\% Wald-type confidence intervals where the standard errors are estimated using nonparametric bootstrap with $1000$ replication.   

It is seen from all of the proposed discriminatory measures that the discriminatory ability of score B was at least as good as score A both in terms of overall risk and $t$-year predicted risk. 

Note that we can also compute the estimate and CI for the difference in discrimination measured for the comparison between scoring rules A and B. This is done in the Supplementary Material.

\section{Concluding remarks}\label{sec:conclucion}
In this paper we proposed novel debiased estimators of the c-index, the concordance probability, and the $t$-year area under the receiver operating characteristic (ROC) curve ($\mbox{AUC}_t$). As demonstrated in our simulation study, the proposed estimators have many desirable properties including (i) do not require data to be generated from a specific model such as the Cox model (ii) robustness to misspecification of the censoring model and (iii) allow for the use of data-adaptive methods for model fitting. We illustrated the use of the estimators by an application to data from a study on brain tumor growth where we compared the discriminatory power of a score (A) which included the tumor growth biomarker FET-VOL and the genetic marker methyl as predictors to a score (B) with  the tumor growth biomarker BV-VOL and the genetic marker methyl as predictors. The analysis showed that the discriminatory ability of score B was at least as good as score A. 

When using the truncated version $K_{\tau}$, the benchmark value is no longer 0.5, which makes it harder to use in practice. However,
if one wishes to compare two scoring rules, $A$ and $B$, then one may calculate the estimate and corresponding confidence interval for the difference in the discrimination measure between rules $A$ and $B$. We illustrate this using the Brain Tumor data in  the Supplementary Material. 

A key step in the development of the proposed estimators was to apply an estimator of the estimand $\beta_t(P)$ that solves (the empirical version) of its corresponding eif equal to zero as we could then avoid estimation of the density function of  the covariate distribution. The specific choice of the estimand $\beta_t(P)$ may be less crucial, however, and others may be pursued as well. We leave this for future work.

The choice of the truncation time $\tau$ may require some care. For the brain tumor data, we took $\tau=2$, which was based on the scientific question at hand. However, in other studies, some choices of $\tau$ may not be feasible, in particular when using bootstrap to estimate standard errors.
There may also be situations in which there is no natural choice of $\tau$. Different practical solutions for choosing $\tau$ have been proposed in the literature; see \cite{cui2023estimating} for an example. We leave a general solution for future research; see \cite{tian2020empirical} for a proposal for a related but simpler problem.

\section*{Acknowledgments}
The authors thank the editors and two anonymous reviewers for their constructive comments, which have led to significant improvement of this paper.

\section*{Conflicts of interest}
None declared.

\section*{Funding}
The PhD project of Marie Skov Breum was funded by a research gift from Novo Nordisk A/S to the Section of Biostatistics, University of Copenhagen.

\section*{Data availability}
The data used in Section \ref{sec:emprical} can be obtained from the authors upon reasonable request.

\begin{appendix} \label{sec:appendix}
\section{}
\subsection{Derivation of the efficient influence function \label{app:eif}}

In this section we derive the efficient influence functions  of $\Psi_\tau(P)$ and $\Theta_t(P)$.
We will start by deriving the eifs in the case without censoring. Let $Z=(T, X)$ denote the `full' data and assume $Z \sim Q$ where $Q$ is a probability distribution belonging to a non-parametric statistical model $\mathcal{Q}$. We also define $N^F(t)=I( T\leq t)$  corresponding to the counting process in the case without censoring and $dM^F(t; X)$ the corresponding martingale increment conditioning on $X$. Further let $dM_{T \mid Y}(t; Y)=dN^F(t) -I(T\geq t)d\Lambda_{T \mid Y}(t \mid Y)$ and note that $dM_{T\mid Y}(t; Y)$ can be written in terms of the martingale increment $dM^F(t;X)$ as follows
\begin{align}
    \label{eq:MGY}
    dM_{T\mid Y}(t; Y) =& dN^F(t) - I(T \geq t)d\Lambda(t \mid X) + I(T \geq t)\left\{d\Lambda(t \mid X) - d\Lambda_{T\mid Y}(t \mid Y) \right\} \nonumber \\
    =& dM^F(t; X) +I(T \geq t)\left\{d\Lambda(t \mid X) - d\Lambda_{T\mid Y}(t \mid Y) \right\} \nonumber \\
    =& dM^F(t ; X) - S(t \mid X) \int_0^t \frac{dM^F(u ; X)}{S(u \mid X)}\left\{d\Lambda(t \mid X) - d\Lambda_{T\mid Y}(t \mid Y) \right\} \nonumber \\
    &+S(t \mid X)\left\{d\Lambda(t \mid X) - d\Lambda_{T\mid Y}(t \mid Y) \right\}, 
\end{align}
where in the last equality we have used that
\begin{align*}
    I(T \geq t) = I(T \geq t) - S(t \mid X) + S(t \mid X) =S(t \mid X) -S(t \mid X) \int_0^t \frac{dM^F(u ;X)}{S(u \mid X)}.
\end{align*}

There are many ways of deriving efficient influence functions. Following \cite{Hines2022demystifying} and \cite{kennedy2023semiparametric} we will derive the eif in the uncensored case by computing the Gâteaux derivative  of the parameter in the direction of a point mass contamination. Specifically we will compute the pathwise derivative
\begin{align*} 
    \frac{d}{d\varepsilon}\bigg\vert_{\varepsilon=0} \Psi(Q_\varepsilon),
\end{align*}
where $Q_{\varepsilon} = (1-\varepsilon)Q + \varepsilon \delta_{\tilde z} $ and $\delta_{\tilde z}$ is the Dirac measure at $Z=\tilde z$.

By \cite{Tsiatis2006} formula (10.76) we can map the eif based on full data $D^{*, F}(Z)$ to the eif based on observed data $D^*(O)$ using the relation 
\begin{align}
\label{eq:Tsiatis}
    	D^*(O) =\frac{\Delta D^{*, F}(Z)}{K_C(\tilde T \wedge t \mid X)}+\int \frac{E\{D^{*, F}(Z)\mid T \geq u, X\}}{{K_C(u \mid X)}} dM_C(u ; X),
\end{align}
where $dM_C(t;X)$ is the martingale (increment) corresponding to the censoring counting process $N_C(t)=I(\tilde T \leq t, \Delta=0)$ conditioning on $X$.

A similar result can also be found in Ex. 1.12 in \cite{laan2003unified}.

Moreover for any function $h(u, X)$ 
\begin{align}
\begin{split}
    \label{eq:Lu}
    \frac{\Delta \int h(u, X) d M^F(u ;X)}{K_{C}(U \mid X)} & +\int \frac{d M_{C}(u ; X)}{K_{C}(u \mid X)} E\left\{\int h(u, X) d M^F(u ; X) \mid T \geq u, X\right\} \\
& =\int \frac{h(v, X)}{K_C(v \mid X)} d M(v ;X).
\end{split}
\end{align}

\subsubsection{Efficient influence function of $\Psi_{\tau}(P)$}
By the chain rule and the Leibniz integral rule we have that the Gâteaux derivative of \eqref{eq:F_Y} is
\begin{align}
    \label{eq:eifFY}
    \frac{d}{d\varepsilon}\bigg\vert_{\varepsilon=0} F_{Y, \varepsilon}(y)=&F_{W\mid V}\left(\frac{y-\beta_V^T \tilde v}{\beta_W}\right)-F_Y(y)+H(\beta;y)\dot{\beta}^F(Q)+I(\beta_V^T\tilde v +\beta_W \tilde w \leq y) \nonumber \\
    &-F_{W\mid V}\left(\frac{y-\beta_V^T \tilde v}{\beta_W}\right) \nonumber \\
    =&I(\tilde y \leq y)-F_Y(y)+H(\beta;y)\dot{\beta}^F(Q),
\end{align}
where 
\begin{equation}
\label{H-const}
    H(\beta;y)=-\frac{1}{\beta_W}\int f_{W\mid V}\left(\frac{y-\beta_V^Tv}{\beta_W}\right)\left [v^T,\frac{y-\beta_V^Tv}{\beta_W}\right ]f_V(v)dv,
\end{equation}
and $\dot{\beta}^F$ is the influence function for $\beta$ based on full data. 

We further note that
\begin{align*}
    \frac{d}{d\varepsilon}\bigg\vert_{\varepsilon=0} \Lambda_{T \mid Y, \varepsilon}(t \mid y) = \frac{\delta_{\tilde y}(y)}{f_Y(y)}\int_0^t \frac{dM_{T \mid Y}(u;y)}{S_{T \mid Y}(u \mid y)},
\end{align*}
and that
\begin{align*}
    \frac{d}{d\varepsilon}\bigg\vert_{\varepsilon=0} S_{T \mid Y, \varepsilon}(t \mid y)= - \frac{\delta_{\tilde y}(y)}{f_Y(y)}S_{T \mid Y}(t \mid y) \int_0^t \frac{dM_{T \mid Y}(u ;y)}{S_{T \mid Y}(u \mid y)},
\end{align*}
so that by the chain rule
\begin{align*}
    \frac{d}{d\varepsilon}\bigg\vert_{\varepsilon=0} h_{\tau, \varepsilon}(y_1, y_2)=& \int_0^\tau \frac{d}{d\varepsilon}\bigg\vert_{\varepsilon=0} S_{T \mid Y, \varepsilon}(t \mid y_2) S_{T\mid Y}(t \mid y_1) d\Lambda_{T\mid Y}(t \mid y_1) \\
    &+\int_0^\tau  S_{T\mid Y}(t \mid y_2)  \frac{d}{d\varepsilon}\bigg\vert_{\varepsilon=0} S_{T \mid Y, \varepsilon}(t \mid y_1)d\Lambda_{T\mid Y}(t \mid y_1)\\
    &+ \int_0^\tau S_{T\mid Y}(t \mid y_2)S_{T\mid Y}(t \mid y_1) \frac{d}{d\varepsilon}\bigg\vert_{\varepsilon=0} d\Lambda_{T \mid Y, \varepsilon}(t \mid y_1) \\
    =&-\frac{\delta_{\tilde y}(y_2)}{f_Y(y_2)}\int_0^\tau S_{T\mid Y}(t \mid y_2) \int_0^t \frac{dM_{T\mid Y}(u ;y_2)}{S_{T\mid Y}(u \mid y_2)}S_{T\mid Y}(t \mid y_1) d\Lambda_{T\mid Y}(t \mid y_1) \\
    &- \frac{\delta_{\tilde y}(y_1)}{f_Y(y_1)}\int_0^\tau S_{T\mid Y}(t \mid y_2)  S_{T\mid Y}(t \mid y_1) \int_0^t \frac{dM_{T\mid Y}(u ; y_1)}{S_{T\mid Y}(u \mid y_1)}d\Lambda_{T\mid Y}(t \mid y_1) \\
    &+ \frac{\delta_{\tilde y}(y_1)}{f_Y(y_1)}\int_0^\tau S_{T\mid Y}(t \mid y_2) dM_{T\mid Y}(t ; y_1). 
\end{align*}
Then
\begin{align*}
    \frac{d}{d\varepsilon}\bigg\vert_{\varepsilon=0} \Psi_{\tau}(Q_{\varepsilon})=&\int \int_{y_1 >y_2} h_\tau(y_1, y_2) dF_Y(y_1) d\frac{d}{d\varepsilon}\bigg\vert_{\varepsilon=0} F_{Y, \varepsilon}(y_2) \\
    &+ \int \int_{y_1 >y_2} h_\tau(y_1, y_2) d \frac{d}{d\varepsilon}\bigg\vert_{\varepsilon=0} F_{Y, \varepsilon}(y_1) dF_Y(y_2) \\
    &+\int \int_{y_1 >y_2} \frac{d}{d\varepsilon}\bigg\vert_{\varepsilon=0} h_{\tau, \varepsilon}(y_1, y_2) dF_Y(y_1)dF_Y(y_2) \\
    =& \int I(y<\tilde y)h_{\tau}(\tilde y,y) dF_Y(y) + \int I(y>\tilde y)h_{\tau}(y,\tilde y) dF_Y(y) +\tilde H(\beta)\dot{\beta}^F(Q) - 2 \Psi_{\tau} \nonumber \\
 &-\int I(y > \tilde y) \int_0^{\tau}S_{T\mid Y}(t\mid \tilde y)\int_0^t\frac{dM_{T\mid Y}(u;\tilde y)}{S_{T\mid Y}(u\mid \tilde y)}S_{T\mid Y}(t\mid y)d\Lambda_{T\mid Y}(t\mid y)dF_Y(y)\nonumber\\
&-\int I(y < \tilde y) \int_0^{\tau}
S_{T\mid Y}(t\mid y)S_{T\mid Y}(t\mid \tilde y)\int_0^t\frac{dM_{T\mid Y}(u;\tilde y)}{S_{T\mid Y}(u\mid  \tilde y)}d\Lambda_{T\mid Y}(t\mid \tilde y)dF_Y(y) \\
&+\int I(y < \tilde y)\int_0^{\tau} S_{T\mid Y}(t\mid y)dM_{T\mid Y}(t;\tilde y)dF_Y(y),
\end{align*}
where
\begin{align}
\label{Constant_H}
    \tilde H(\beta)= \int \int_{y_1>y_2} h_\tau(y_1, y_2) dF_Y(y_1)dH(\beta; y_2)  +  \int \int_{y_1>y_2} h_\tau(y_1, y_2) dH(\beta; y_1)dF_Y(y_2) ,
\end{align}
with $H(\beta; y)$ defined in \eqref{H-const}.

Hence the eif based on full data is 
\begin{align*}
    D^{*, F}_{\Psi_\tau}(Z)(Q)=& \int I(y<Y)h_{\tau}(Y,y) dF_Y(y) + \int I(y>Y)h_{\tau}(y,Y) dF_Y(y) +\tilde H(\beta)\dot{\beta}^F(Q) - 2 \Psi_{\tau}\nonumber \\
 &-\int I(y > Y) \int_0^{\tau}S_{T\mid Y}(t\mid Y)\int_0^t\frac{dM_{T\mid Y}(u;Y)}{S_{T\mid Y}(u\mid Y)}S_{T\mid Y}(t\mid y)d\Lambda_{T\mid Y}(t\mid y)dF_Y(y)\nonumber\\
&-\int I(y < Y)\int_0^{\tau}
S_{T\mid Y}(t\mid y)S_{T\mid Y}(t\mid Y)\int_0^t\frac{dM_{T\mid Y}(u;Y)}{S_{T\mid Y}(u\mid  Y)}d\Lambda_{T\mid Y}(t\mid Y)dF_Y(y) \\
&+\int I(y < Y)\int_0^{\tau} S_{T\mid Y}(t\mid y)dM_{T\mid Y}(t;Y)dF_Y(y).
\end{align*}

Combining \eqref{eq:MGY},  \eqref{eq:Tsiatis}  and \eqref{eq:Lu} we can write the eif based on observed data as in Theorem \ref{theorem:eif1}.

\subsubsection{Efficient influence function of $\Theta_t(P)$}
Note that
\begin{align*}
     \frac{d}{d\varepsilon}\bigg\vert_{\varepsilon=0} v_{t,\varepsilon}(y_1, y_2)=& -\frac{d}{d\varepsilon}\bigg\vert_{\varepsilon=0} S_{T \mid Y, \varepsilon}(t \mid y_1) S_{T \mid Y}(t \mid y_2) + \left\{1- S_{T \mid Y}(t \mid y_1) \right\}\frac{d}{d\varepsilon}\bigg\vert_{\varepsilon=0} S_{T \mid Y, \varepsilon}(t \mid y_2) \\
     =& \frac{\delta_{\tilde y}(y_1)}{f_Y(y_1)}S_{T \mid Y}(t \mid y_1) \int_0^t \frac{dM_{T \mid Y}(u ; y_1)}{S_{T \mid Y}(u \mid y_1)}S_{T \mid Y}(t \mid y_2) 
     \\&-\left\{1- S_{T \mid Y}(t \mid y_1) \right\}\frac{\delta_{\tilde y}(y_2)}{f_Y(y_2)}S_{T \mid Y}(t \mid y_2) \int_0^t \frac{dM_{T \mid Y}(u; y_2)}{S_{T \mid Y}(u \mid y_2)}.
\end{align*}
Then by the chain rule and \eqref{eq:eifFY} we have
\begin{align*}
    \frac{d}{d\varepsilon}\bigg\vert_{\varepsilon=0} \Theta_t(Q_{\varepsilon}) =& \int I(y<\tilde y)v_t(\tilde y, y) dF_Y(y) 
    + \int I(y>\tilde y)v_t(y, \tilde y) dF_Y(y) - 2\Theta_t\\
    &+\int I(y < \tilde y)S_{T \mid Y}(t \mid y)S_{T \mid Y}(t \mid \tilde y) \int_0^t\frac{dM_{T \mid Y}(u ;\tilde y)}{S_{T \mid Y}(u \mid \tilde y)}dF_Y(y) \\
    &-\int I(y > \tilde y) v_t(y, \tilde y)\int_0^t\frac{dM_{T \mid Y}(u ;\tilde y)}{S_{T \mid Y}(u \mid \tilde y)}dF_Y(y)+\bar{H}(\beta)\dot{\beta}^F(Q)  \\
    =& \int I(y<\tilde y)v_t(\tilde y, y) dF_Y(y) 
    + \int I(y>\tilde y)v_t(y, \tilde y) dF_Y(y)  - 2\Theta_t\\
    &+ \left\{ F_T(t) - F_Y(\tilde y) \right\}S_{T \mid Y}(t \mid \tilde y)\int_0^t\frac{dM_{T \mid Y}(u ;\tilde y)}{S_{T \mid Y}(u \mid \tilde y)} +\bar{H}(\beta)\dot{\beta}^F(Q) ,
\end{align*}
where
\begin{align}
\label{eq:Constant_H_AUC}
    \check{H}(\beta)= \int \int_{y_1>y_2} v_t(y_1, y_2) dH(\beta; y_1) dF_Y(y_2) +  \int \int_{y_1>y_2} v_t(y_1, y_2) dF_Y(y_1) dH(\beta; y_2)  ,
\end{align} 
with $H(\beta; y)$ defined in \eqref{H-const}.
Hence the eif under full data is
\begin{align*}
D_{\Theta_t}^{*,F}(Z)(Q) =&  \int I(y<Y)v_t(Y, y) dF_Y(y) + \int I(y>Y)v_t(y, Y) dF_Y(y) - 2\Theta_t\nonumber \\
    &+ \left\{ F_T(t) - F_Y(Y) \right\}S_{T \mid Y}(t \mid Y)\int_0^t\frac{dM_{T \mid Y}(u ; Y)}{S_{T \mid Y}(u \mid Y)} +\bar{H}(\beta)\dot{\beta}^F(Q) .
\end{align*}

Combining \eqref{eq:MGY},  \eqref{eq:Tsiatis}  and \eqref{eq:Lu} we can write the eif based on observed data as in Theorem \ref{theorem:eif2}.

\subsection{Second order remainder term and robustness properties}\label{app:remainder}

We focus on the estimand $\Psi_{\tau}$, and the goal is to investigate the robustness properties
 of the proposed estimator of $\Psi_{\tau}$
in terms of consistency as we have already pointed out that we
will not use the influence function for inference anyway. 
In order to study the robustness properties of $\Psi_{\tau}$ we consider the von Mises expansion
\begin{align}
    \Psi_\tau(P_n) - \Psi_\tau(P)  =& - E \left\{ D^*_{\Psi_\tau}(O)(P_n) \right\}  + R_n^{\Psi_{\tau}} \nonumber \\
    \label{eq:term1}
    =& \mathbb{P}_n D^*_{\Psi_\tau}(O)(P) \\
    \label{eq:term2}
    &- \mathbb{P}_n D^*_{\Psi_\tau}(O)(P_n) \\
    \label{eq:term3}
    &+ (\mathbb{P}_n-P) \left\{D^*_{\Psi_\tau}(O)(P_n)- D^*_{\Psi_\tau}(O)(P) \right\} \\
    \label{eq:term4}
    &+R_n^{\Psi_{\tau}},
\end{align}
where $R_n^{\Psi_{\tau}}= E \left\{ D^*_{\Psi_\tau}(O)(P_n) \right\} + \Psi_\tau(P_n) - \Psi_\tau(P)$. 

The term in \eqref{eq:term1} is asymptotically normally distributed by the Central Limit Theorem. The term in \eqref{eq:term2} is the so-called `plug-in bias' or first order bias term. The one-step estimator $\Psi_\tau(P_n) + \frac{1}{n} \sum_{i=1}^n D^*_{\Psi_\tau}(O_i)(P_n)$ accounts for this bias by construction. 
We need both the empirical process term in \eqref{eq:term3}  and the second order remainder in \eqref{eq:term4}  to be $o_p(1)$ for the one-step estimator to be consistent. The empirical process term can be controlled by either Donsker class conditions \citep{vanderVaart2000} or sample splitting \citep{Chernozhukov2018}. We analyze the second order remainder term further below. However, we  study first the large sample properties of the proposed estimator of $\beta_t(P)$.

\subsubsection{Remainder term of of $\beta^{AL}_t(P)$}\label{app:beta}
Recall the definitions $\beta_1=\mbox{cov}[g\{S(t\mid X)\},X]$ and $\beta_2= \mbox{var}(X)$. From now on we consider the setting where the link function is $g(x)=\log(-\log(x))$. 

We first consider the remainder term
$$
R_n^{\beta_1}=E\{D^*_{\beta_1}(O,P_n)\}+\beta_{1n}-\beta_1,
$$
which can be calculated to
\begin{align*}
R_n^{\beta_1}
=&-E \left\{ (X-\mathbb{P}_nX)\left\{S_n(t \mid X) \Lambda_n(t \mid X)\right\}^{-1} \left\{S_n(u \mid X)-S(u \mid X) \right\}\right\}\\
&-E \left\{ (X-\mathbb{P}_nX)\left\{\Lambda_n(t \mid X)\right\}^{-1}\int_0^t\frac{S(u \mid X)}{S_n(u\mid X)} \frac{ K_C(u \mid X) - K_{C,n}(u \mid X) }{K_{C,n}(u\mid X)} \left\{d\Lambda(u \mid X) - d\Lambda_n(u \mid X) \right\} \right\}\\
&+ E \Big(\Big[\log\{\Lambda_n(t\mid X)\}-\log\{\Lambda(t\mid X)\}\Big](X-\mathbb{P}_nX)\Big)\\
&+ \Big[E\log\{\Lambda(t\mid X)\}-\mathbb{P}_n\log\{\Lambda_n(t\mid X)\}\Big](EX-\mathbb{P}_nX).
\end{align*}

Hence, if we use estimators $K_n(u\mid X)$ and $S_n(u\mid X)$ that converges to the  truth sufficiently fast (faster than rate $n^{1/4}$), $n^{1/2}R_n^{\beta_1}$ converges to zero in probability. It is also easily seen that 
$$
n^{1/2}R_n^{\beta_2}=n^{1/2}(EX-\mathbb{P}_nX)(EX-\mathbb{P}_nX)^T,
$$
that converges to zero in probability. We then have
$$
D^*_{\beta}(O,P)=\beta_2^{-1}D^{*}_{\beta_1}(O,P)-\beta_2^{-1}D^{*}_{\beta_2}(O,P)\beta,
$$
which is then  also the influence function of $\beta_n$ if estimators that converges to the  truth faster than rate $n^{1/4}$ are used. 
This also means that 
$$
n^{1/2}E\{D^*_{\beta}(O,P))\}=-n^{1/2}(\beta_n-\beta)+o_P(1),
$$
as $n^{1/2}R_n^{\beta}$ converges to zero in probability.

\subsubsection{Remainder term of $\Psi_\tau(P)$}

From now on we assume that sufficiently fast converging estimators have been used in the estimation process for $\beta$ so that consistency is guaranteed. Because we will only investigate the 
robustness properties
 of the proposed estimator of $\Psi_{\tau}$
in terms of consistency we can leave out terms that contains the contrast $F_Y^n-F_Y$ (for fixed $\beta$) as these terms are $o_p(1)$. It is instructive to start deriving the second order remainder term in the case of full data (ie no censoring). For fixed $\beta$ (ie ignoring the eif of $\beta$) we get after some algebra that 
\begin{align*}
R_n^{\Psi_{\tau}}=-\int\int_{y_1>y_2} \{S_{T \mid Y,n}(t\mid y_2)-S_{T \mid Y}(t\mid y_2)\}   d\{F_{T\mid Y}^n(t\mid y_1)-F_{T \mid Y}(t\mid y_1)\}dF_Y(y_1)dF_Y(y_2)+o_p(1),
\end{align*}
where we define $dF_{T\mid Y}(t\mid y)=S_{T \mid Y}(t\mid y)d\Lambda_{T \mid Y}(t\mid y)$.

\noindent
Define further 
\begin{align*}
    dG_n(u\mid X)&=\biggl(\frac{K_C^n-K_C}{K_C^n}\biggr )(u \mid X)\{d\Lambda_n(u \mid X)-d\Lambda(u \mid X)\}\\
    dR_n(u;X,Y)&=d\Lambda_n(u \mid X)-d\Lambda_{T \mid Y,n}(u \mid Y).
\end{align*}

\noindent
In the observed data case we get the same term as in the full data case plus some additional product terms involving the estimated censoring survival function. Specifically we get after some further algebra that
\footnotesize
\begin{align*}
R_n^{\Psi_{\tau}}=&-\int\int_{y_1>y_2} \int_0^{\tau}\{S_{T \mid Y, n}(t\mid y_2)-S_{T \mid Y}(t\mid y_2)\}   
d\{F_{T \mid Y}^n(t\mid y_1)-F_{T \mid Y}(t\mid y_1)\}dF_Y(y_1)dF_Y(y_2)\\
&+E\left\{\int I(y<Y)\int_0^{\tau}S_{T \mid Y,n}(t\mid y)S(t \mid X)dG_n(t \mid X)dF_Y(y)\right\}\\
&-E\left\{\int I(y<Y)\int_0^{\tau}S_{T \mid Y,n}(t\mid y)S(t\mid X)
\int_0^t\frac{S(u\mid X)}{S_n(u\mid X)}
dG_n(u\mid X)
dR_n(t;X,Y)
dF_Y(y)\right\}\\
&+E\left\{\int I(y<Y)\int_0^{\tau}S_{T \mid Y,n}(t\mid y)
\int_0^t\frac{S(u\mid X)}{S_n(u\mid X)}
dG_n(u\mid X)F_{T \mid Y}^n(t \mid y) dF_Y(y)\right\}\\
&-E\left\{\int I(y<Y)\int_0^{\tau}S_{T \mid Y}n(t\mid y)
\int_0^t\frac{S(u\mid X)}{S_{T \mid Y,n}(u\mid Y)}\int_0^{u}\frac{S(v\mid X)}{S_n(v\mid X)}
dG_n(v\mid X)dR_n(u;X,Y)F_{T \mid Y}^n(t \mid y) dF_Y(y)\right\}\\
&-E\left\{\int I(y>Y)\int_0^{\tau}S_{T \mid Y,n}(t\mid Y)
\int_0^t\frac{S(u\mid X)}{S_n(u\mid X)}
dG_n(u\mid X)F_{T \mid Y}^n(t \mid y) dF_Y(y)\right\}\\
&+E\left\{\int I(y>Y)\int_0^{\tau}S_{T \mid Y,n}(t\mid Y)
\int_0^t\frac{S(u\mid X)}{S_{T \mid Y,n}(u\mid Y)}\int_0^{u}\frac{S(v\mid X)}{S_n(v\mid X)}
dG_n(v\mid X)dR_n(u;X,Y)F_{T \mid Y}^n(t \mid y) dF_Y(y)\right\}\\
&+o_p(1),
\end{align*}
\normalsize
and if the needed empirical quantities are estimated using an estimator that converges at faster rate than $n^{1/4}$ the remainder term will go to zero in probability, and we get the EIF as the IF. So in this respect the proposed estimator is efficient. However, we do not recommend to use the EIF to estimate the variability of the proposed estimator as one would then need to estimate the complicated constant $\tilde H$.

\subsection{Bias correction}\label{app:bias}
We wish to highlight the subtle bias correction achieved by our method, exemplified using the estimand targeted by \cite{GonenHeller2005} (GH): $2\Psi_{\tau}(P)$, where 
\begin{align*}
\Psi_{\tau}(P)=\int\int_{y_1>y_2}  h_{\tau}(y_1,y_2)dF_Y(y_1)dF_Y(y_2),
\end{align*}
with $F_Y$ the distribution function for $Y$, and 
\begin{align}
\label{eq:h}
h_{\tau}(y_1,y_2)=\int_0^{\tau} S_{T \mid Y}(t |  y_2)S_{T \mid Y}(t |  y_1)d\Lambda_{T \mid Y}(t |  y_1).
\end{align}
Denote also $S_{T \mid Y}(t|y_1)d\Lambda_{T \mid Y}(t|y_1)$ by $f_{T \mid Y}(t|y_1)dt$ and
let $h_{\tau}^n$ denote the above $h_{\tau}$ with $S$ and $\Lambda$ replaced by $S_n$ and $\Lambda_n$. The GH-estimator is a plug-in estimator using a $h_{\tau}^n$ based on Cox regression (and using the emiprical distribtuion function $F_n$ for $F$) in $\Psi_{\tau}(P)$.
Clearly, this procedure is biased if the Cox model is incorrectly specified as $h_{\tau}^n$ then converges (in probability) to something different from $h_{\tau}$. 
Specifically, the bias is (up to second order) given by:
\begin{align*}
 &\int_0^{\tau}\{S_{T \mid Y,n}(t |  y_2)-S_{T \mid Y}(t |  y_2)\}f_{T \mid Y}(t|y_1)dt+\int_0^{\tau} S_{T \mid Y}(t |  y_2)\{f_{T \mid Y,n}(t|y_1)-f_{T \mid Y}(t|y_1)\}dt   \\
 +&\int_0^{\tau}\{S_{T \mid Y,n}(t |  y_2)-S_{T \mid Y}(t |  y_2)\}\{f_{T \mid Y,n}(t|y_1)-f_{T \mid Y}(t \mid y_1)\}dt
\end{align*}
(these term are further integrated over $(y_1,y_2)$  wrt $I(y_1>y_2)dF_y^n(y_1)dF_y^n(y_2)$).
Our estimator, using  a working Cox model, removes the first order bias terms (first two terms in the latter display), which is why it can  still  show some bias, although it is a second order bias term corresponding to the last term in the latter display.
However, if we use machine learning techniques instead, then the second order bias term is removed as well. There are some further bias terms due to estimation of the censoring  distribution, but these vanishes as well if machine learning is used at this stage as well.
We believe this nicely shows the benefit of using the efficient influence function as the starting point for developing estimation.

\end{appendix}

\bibliographystyle{plainnat} 
\bibliography{ref.bib}


\end{document}